\documentclass[a4paper,11pt]{article}
\usepackage[dvipdfmx,hiresbb]{graphicx}

\usepackage{jheppub} 
                     
\usepackage[T1]{fontenc} 

\title{\boldmath 
Green's functions in the presence of a bubble wall
}

\author[a]{Takahiro Kubota  
}

\affiliation[a]{CELAS, Osaka University, Toyonaka, Osaka, 560-0043, Japan}

\emailAdd{takahirokubota859@hotmail.com}

\abstract{
Field theoretical tools  are developed so that one can  analyze quantum phenomena such as transition radiation that must have occurred  during the Higgs condensate  bubble expansion through plasma in the early universe.   Integral representations of Bosonic and Fermionic propagators are presented  for the case that particle masses are varied continuously during the passage through the bubble wall  interface between symmetry-restored and symmetry-broken regions.   The construction of propagators is based on the so-called eigenfunction expansion method associated  with self-adjoint  differential operators,  developed by Weyl, Stone, Titchmarsh,  Kodaira  and several others.  A novel method of field quantization in the presence of the  bubble wall is proposed by using the spectral functions introduced in constructing the two-point Green's functions. 
}

\begin{document} 
\maketitle
\flushbottom

\section{Introduction}
\label{sec:intro}
The discoveries of the gravitational waves  by LIGO and Virgo collaboration  
 \cite{abbott0} - \cite{abbott3}
are apparently a strong driving force, urging us  not only to look for further astrophysical sources but also to gain a direct access to the physics of the early universe by observing gravitational radiation \cite{witten} - \cite{kosowsky2} .
The  standard model  phase transition is smooth cross-over \cite{kajantie} - \cite{olivergould},
but extension of the electroweak  model can lead to first-order  phase transitions 
 \cite{anderson} - \cite{grojean}
 that could  have produced observable gravitational waves in accessible frequency ranges for future experiments.  The LISA space-based interferometer  \cite{caprini1} - \cite{auclair},   which is   planned to have sensitivity band peaks at lower frequencies than ground based detectors, is expected  to probe first-order  phase transitions of the extended electroweak models in the early universe. Other planned space-based gravitational wave observation projects include DECIGO \cite{kawamura}, Taiji \cite{ruan} and TianQin \cite{luo}.

The strength and spectral shape of the gravitational wave  in the phase transitions depend on   several parameters, which include  (i)  the characteristic time duration   of the phase transition,  (ii) the ratio of the vacuum energy density released in the transition to  the radiation bath,  (iii) the temperature and the Hubble parameter  at the time of the phase transition,  (iv) the fraction of vacuum energy converted  into bulk motion of the fluid and into the gradient energy of the Higgs field  and  (v) the bubble wall velocity in the plasma rest frame   \cite{caprini1} - \cite{auclair}. 

Among these parameters the bubble wall velocity is the least clearly understood quantity. Unless the microscopic mechanism of friction exerted by the fluid is  fully understood, the wall velocity is basically left  as an ambiguous  parameter in the hydrodynamical treatment of the plasma. The wall velocity  is also an important quantity  for the generation of the matter-antimatter asymmetry 
\cite{kuzmin} - \cite{garbrecht}.   B\"odeker and Moore \cite{boedeker1} have once investigated the propagation speed of the interface  between symmetry-restored and symmetry-broken regions.  Particles crossing the bubble wall lose some part of their momentum in the direction perpendicular  to the wall, thereby impeding the expansion of the bubble wall. The question was whether the bubble wall would  become relativistic or even ultra-relativistic and therefore run away.  They extended the standard electroweak theory with extra scalar fields and looked for a simple criterion to determine  whether or not the runaway of the bubble wall could  occur.  Espinosa et al  \cite{espinosa1} also made a comprehensive study of the bubble wall velocity, classifying the bubble expansion regimes into deflagrations, detonations, hybrids and runaway.

Several years later,  B\"odeker and Moore  revisited the bubble wall runaway problem  \cite{boedeker2},   by considering  transition radiation effects \cite{ginzburgfrank, jackson}, which had been  neglected in their previous work   \cite{boedeker1}.  When particles hit the bubble wall, they may well radiate particles with phase-dependent mass, such as $W$- and $Z$-bosons, and the emission induces  important effects on the bubble wall. It was argued that the force is dominated by soft particle radiation  and that  the friction force could be significantly altered due to logarithmically-enhanced radiative corrections. The appearance of the large logarithms is traced back to the  soft momentum region in the phase space  and is therefore of infrared character. The analyses in \cite{boedeker2}  showed that without considering the transition radiation the friction on the wall would be underestimated. 

The transition radiation considered in   \cite{boedeker2} is due to the process of a particle decaying into two, one of which is a soft vector boson.  If such a process is important, then a question naturally arises whether the processes of a particle decaying into several soft particles could be equally important. Fixed order calculations will quite probably break down for very large wall velocity. H{\" o}che et al \cite{hoeche} therefore attempted to perform the all-order summation  of the large logarithmic terms due to the multiple emission of soft gauge bosons. In the course of the calculation they also payed attention to double logarithms of the Sudakov type \cite{sudakov}, whose all-order resummation and exponentiation had been  extensively studied  in QED \cite{mueller, collins3} and  QCD \cite{sen1} - \cite{contopanagos}. 
After all they concluded that the logarithmic enhancement implies significantly slower terminal velocities than previously thought, to find a peculiar scaling law of the pressure.  (See also \cite{marc} for similar scaling law.)

Their formulation in \cite{hoeche} reminds us of the parton shower algorithm \cite{webber, buckley} in hadron collisions formulated as an evolution in momentum transfer from high energy scale to the lower.  In the QCD-based resummation technique, the experimental resolution scale plays the role of regulator of infrared singularities. In the electroweak radiative corrections, the dominant contributions are also expected to come from  double logarithms of the Sudakov type as well as single logarithms involving $\sqrt{s}/{M_{Z}}^{}$, when the energy  $\sqrt{s}$  of the system is much larger than  the electroweak energy scale, say, the $Z$-boson mass $M_{Z}$ ( i.e., $\sqrt{s} \gg M_{Z}$).   The origin of these logarithms is in close correspondence with the soft and collinear singularities known well in massless gauge theories.
In contrast to the massless gauge theories such as QED and QCD where cancellation occurs among virtual corrections  and real gauge boson radiation,  the masses of $Z$- and $W$-bosons in the electroweak theory  provide   the physical cutoff and the large logarithms  originating from virtual corrections  are of direct physical significance \cite{dennerpozzorini1}- \cite{lindert}.

 The processes that must have occurred during the bubble expansion in the early universe  are somewhere between massless and massive gauge boson  theories, and we have to take care of the fact that  the masses of the standard model particles, irrespectively of real or virtual ones,  may change near the bubble wall.  In Ref.   \cite{hoeche}  real emissions of gauge bosons and virtual corrections were combined together to see infrared divergence cancellations, while neglecting the change of masses of vector bosons at the domain wall. It order to see the varying mass effects, it might be tempting to use the so-called vacuum expectation value insertion approximation, but we must exert great care in using such an approximation  when infrared singularities are involved. 
 
 In the last several years a lot of efforts have been made  \cite{laurent1} -  \cite{azatov4} in order to understand in a better way the dynamics occurring between plasma particles and bubble wall.  We must say, however,  that we are not yet  fully equipped with theoretical tools to perform quantum field theoretical analyses such as loop calculations.  One of the sources of difficulties is that we do not have at hand a suitable way   of constructing  field propagators in which the bubble  wall background effects are   fully taken into account. This unsatisfactory situation is closely connected with the lack of suitable field quantization methods for the case in which particles masses change between  symmetry-restored and symmetry-broken regions.  
 
 The purpose of the present paper is to improve  the present status quo  by constructing explicitly the two-point Green's functions, i.e., propagators of scalar, spinor and vector fields for the case that particle masses are position-dependent.   Throughout the present paper we will work in the Lorentz frame in which the bubble wall is at rest.  We will solve the partial differential equations for  the propagators  with position-dependent mass   by making use of the so-called eigenfunction expansion method associated with self-adjoint differential operators  developed  several decades  ago  in mathematical literatures \cite{weyl} - \cite{kodaira2}. Our method is general enough and is applicable irrespectively of the details  of the shape of the bubble wall.  Our two-point Green's functions are expressed in the form of integral representations over the spectrum described by    spectral functions.  On the basis of the spectral functions, the field quantization method is also reexamined    for the case that masses are position-dependent. We will introduce   operators that are analogous to the ordinary creation and annihilation operators of particles and will give the definition of the vacuum using these operators.   It is argued that the propagators to be derived in this work are the expectation value  of time-ordered product of fields with respect to this vacuum. 
 
The  first-principle method of field quantization in the presence of a bubble wall was recently  discussed 
in detail by Azatov et al  \cite{azatov4}. We  agree  completely with their  view on the necessity of the first-principle field-quantization method, while  the approach taken in the present paper differs from theirs in that  emphasis is placed on  the roles played by the spectral functions to be discussed below.

\section{The scalar particle's  two-point Green's function}
\label{sec:green}

\subsection{The method {\` a} la  Weyl, Stone, Titchmarsh and Kodaira}
\label{subsec:wstc}

The method of constructing the propagators  in the presence of the bubble wall  is most easily explained for the case of the spin zero particle obeying the Klein-Gordon equation.     We suppose  that the bubble wall is so large in comparison with our microscopic length scale and that the front of the wall can be approximated by a nearly  planar wall. For the sake of simplicity we assume that the wall is spreading infinitely along the $x-y$-plane. This means that  the expectation value of the Higgs field as well as particle masses depend on the $z$-coordinate only.  Denoting the mass by  $M(z) $, we identify $M(+\infty)$ and $M(-\infty )$ with the mass in the symmetry-broken and unbroken regions, respectively.  For spineless particles satisfying the Klein-Gordon equation, we would like to look for  the Green's function that   satisfies  
\begin{eqnarray}
& & 
\left \{
\square + M(z)^{2}
\right \}
{\cal G} \left ( t-t^{\: \prime}, x-x^{\: \prime}, y-y^{\: \prime}, z, z^{\: \prime} ; M(*) \right ) 
  \nonumber \\
  & & \hskip3cm
= -i \: \delta(t-t^{\prime})\delta(x-x^{\: \prime}) \delta(y-y^{\: \prime})\delta(z-z^{\: \prime})\:,
\label{eq:greensfunction}
\\
& & \left \{
\square^{\: \prime} + M(z^{\: \prime})^{2}
\right \}
{\cal G} \left ( t-t^{\: \prime}, x-x^{\: \prime}, y-y^{\: \prime}, z, z^{\: \prime} ; M(*) \right ) 
\nonumber \\
& & \hskip3cm
= -i \: \delta(t-t^{\prime})\delta(x-x^{\: \prime}) \delta(y-y^{\: \prime})\delta(z-z^{\: \prime})\:,
\label{eq:greensfunction2}
\end{eqnarray}
where $\square$  and $\square ^{\: \prime}$ are the usual d'Alembertian operators
\begin{eqnarray}
\square =\frac{\partial^{2}}{\partial t^{2}}-\frac{\partial^{2}}{\partial x^{2}}-\frac{\partial^{2}}{\partial y^{2}}-\frac{\partial^{2}}{\partial z^{2}}\:, 
\hskip1cm 
\square^{\:\prime}  =\frac{\partial^{ 2}}{\partial t^{\:
 \prime \: 2}}-\frac{\partial^{2}}{\partial x^{\:\prime \: 2}}
 -\frac{\partial^{2}}{\partial y^{\: \prime \: 2}}-\frac{\partial^{2}}{\partial z^{\: \prime \: 2}}\:.
\end{eqnarray}
Our notation $M(*)$ as opposed to  $M(z)$ or $M(z^{\:\prime})$  in (\ref{eq:greensfunction})  and   (\ref{eq:greensfunction2})  is meant that  the Green's function depends only indirectly on the functional form of the  $z$-dependent mass.   Note that the translational invariances in the $t$-,  $x$- and $y$- directions are just as usual and one can express the solution of  (\ref{eq:greensfunction})   and   (\ref{eq:greensfunction2})    as a superposition of  the plane waves in the $x$- and $y$-directions.  The translational invariance in the $z$-direction, however,  is not respected due to the presence of the wall and the mass term $M(z)^{2}$, and  therefore  (\ref{eq:greensfunction2})   does not follow from (\ref{eq:greensfunction}) immediately. 

We now  argue  that a certain mathematical framework developed    in Refs.   \cite{weyl} - \cite{kodaira2}
 is quite useful to  give a closed form of the solutions to   (\ref{eq:greensfunction}) and (\ref{eq:greensfunction2}).   It is a generalized  expansion method associated with second-order differential equations and  is applicable irrespectively of the detailed form of the function $M(z)$.  We will look for the solution to (\ref{eq:greensfunction}) and   (\ref{eq:greensfunction2})  in the following   integral representation
\begin{eqnarray}
& & \hskip-1.5cm 
{\cal G}  \left ( t-t^{\prime}, x-x^{\prime}, y-y^{\prime}, z, z^{\prime} ; M(*)  \right ) 
\nonumber \\
&=&
i \int \frac{dE\: d^{2}\vec{p}_{\perp}}{(2\pi)^{3}}
e^{-iE(t-t^{\prime})}e^{i \vec{p}_{\perp} \cdot ( \vec{x}_{\perp}  - \vec{x}_{\perp} ^{\: \prime})}
\int \sum _{k, l=1, 2}^{}
\frac{
\phi _{k} (z; \lambda )  d\rho_{kl} (\lambda)_{}    \phi_{l} (z^{\prime} ; \lambda )^{}
}
{E^{2} - {\vec{p}_{\perp}}^{\: 2} - \lambda + i \varepsilon  }\;, 
\label{eq:integralrepresentation}
\end{eqnarray}
where $\phi_{k}(z;\lambda) $ is the solution to the following differential equation  in the infinite region $ - \infty < z < +\infty$
\begin{eqnarray}
\left \{
-\frac{d^{2}}{dz^{2}} + M(z)^{2}
\right \}\phi_{k}   (z; \lambda  )
= \lambda \: \phi_{k}  (z; \lambda)\:, 
\:\:\:\:(k = 1, 2)
\label{eq:eigenvalueproblem}
\end{eqnarray}
and $\lambda $ is a parameter of the theory.   We have introduced in (\ref{eq:integralrepresentation})  the two-dimensional position and momentum vectors, 
$\vec{x}_{\perp}=(x, y, 0)$,  $\vec{x}_{\perp}^{\:\prime}=(x^{\:\prime}, y^{\:\prime}, 0)$ and  $\vec{p}_{\perp}=(p_{x},p_{y}, 0)$  together with  the two-dimensional integration volume $d^{2}\vec{p}_{\perp}=dp_{x}\:dp_{y}$.  

We notice immediately  that  Eq. (\ref{eq:eigenvalueproblem}) is the one-dimensional Schr{\" o}dinger-type equation with  ``potential term''  $M(z)^{2}$ and  the ``energy eigenvalue''  $\lambda$.  Since this  is the second-order differential equation, we have two independent solutions,  which we denote by $\phi_{1}(z; \lambda)$ and $\phi_{2}(z; \lambda)$, respectively, and will be called fundamental solutions. The normalization of these solutions is fixed by the initial conditions at $z=0$, i.e., 
\begin{eqnarray}
\phi_{1}(0;\lambda)&=&1, \hskip1cm \phi_{1}^{\: \prime}(0;\lambda )=0 \:,
\label{eq:initialconditions1}
\\
\phi_{2}(0;\lambda)&=&0, \hskip1cm \phi_{1}^{\: \prime}(0;\lambda )=1\:. 
\label{eq:initialconditions}
\end{eqnarray}
The derivative with respect to $z$ is denoted by prime ($\prime$) in (\ref{eq:initialconditions1}) and (\ref{eq:initialconditions}).   The spectrum  of  $\lambda$ is described by $2 \times 2$  spectral function matrix  $\rho_{kl} (\lambda)_{}$  and Eq. (\ref{eq:integralrepresentation}) is the Stieltjes    integral   with respect to $\rho _{kl} (\lambda )_{}$\:. This spectral function will play the most important roles throughout the present work.  Since we are considering Eq. (\ref{eq:eigenvalueproblem}) in the infinite region  $- \infty < z < +\infty $, our spectral  problem is more general and  is more singular in a sense than  the Sturm-Liouville boundary value problem, in which we usually  deal with finite regions.  As we will see later, (\ref{eq:integralrepresentation}) should go over  to the conventional  Feynman's propagator if the mass is $z$-independent, and we have added $+ i\varepsilon $  by hand in the denominator of the integrand of (\ref{eq:integralrepresentation}).  This $+i \varepsilon $ prescription indicates that the Green's function (\ref{eq:integralrepresentation})  corresponds to the vacuum expectation value of the time-ordered product of relevant scalar fields, but the justification of the prescription will be given only after we specify what we mean by the vacuum in the presence of the bubble wall.

Let us apply the differential operator $\left \{  \square + M(z)^{2} \right \}$ on both  sides of (\ref{eq:integralrepresentation}) and we then  obtain 
\begin{eqnarray}
& & \hskip-2cm
\left \{
\square + M(z)^{2}
\right \}
{\cal G}  \left ( t-t^{\prime}, x-x^{\prime}, y-y^{\prime}, z, z^{\prime} ; M(*)  \right ) 
\nonumber \\
& = &
i \int \frac{dE\:d^{2}\vec{p}_{\perp}\:  }{(2\pi)^{3}}
e^{- iE(t-t^{\prime})}e^{i \vec{p}_{\perp} \cdot (\vec{x}_{\perp} - \vec{x}_{\perp} ^{\:\prime})}
\nonumber \\
& & \times 
\int \sum_{k,  l =1, 2}^{}  \left \{
-E^{2}+\vec{p}_{\perp}^{\:2}   -  \frac{\partial^{2}}{\partial z^{2}} + M(z)^{2}
\right \}
\frac{
\phi_{k} (z; \lambda)
d \rho _{kl} ( \lambda)_{} 
\phi_{l}  (z^{\prime}; \lambda)^{}
}
{E^{2} - \vec{p}_{\perp}^{\:2} - \lambda  + i \varepsilon }\:.
\label{eq:applydiffoperation}
\end{eqnarray}
Using the differential  equation (\ref{eq:eigenvalueproblem})  in (\ref{eq:applydiffoperation}) we immediately arrive at 
\begin{eqnarray}
& & \hskip-2cm
\left \{
\square +  M(z)^{2}
\right \}
{\cal G} \left ( t-t^{\prime}, x-x^{\prime}, y-y^{\prime}, z, z^{\prime} ; M(*)  \right ) 
\nonumber \\
& = &
-i \: \delta (t-t^{\prime})\delta (x-x^{\prime})\delta (y-y^{\prime}) 
\int 
\sum_{k, l =1, 2}
\phi_{k}  (z; \lambda ) d \rho _{kl} (\lambda) _{} \phi_{l}  (z^{\prime}; \lambda)^{}\:.
\label{eq:beforeusingcompleteness}
\end{eqnarray}
By virtue of   the work of Weyl  \cite{weyl} and  of Stone \cite{stone} it has been known  that there exists  the matrix $\rho_{kl} (\lambda )$ with  such a nice property  that any  real-valued  {\it square integrable} function  $f(z)$ can be expressed as a superposition of the fundamental solutions 
$\phi_{k}(z; \lambda)\:\:\:(k=1,2)$, i.e., 
\begin{eqnarray}
f(z)=\int \sum_{k, l =1,2}  \phi_{k}(z; \lambda) d\rho_{kl}(\lambda)\int _{-\infty}^{\infty} 
dz^{\: \prime}   \phi_{l}(z^{\:\prime} ; \lambda)   f( z^{\;\prime} )\:.
\label{eq:weylstone}
\end{eqnarray}
In other words we are able to write down   the formula 
\begin{eqnarray}
\int 
\sum_{k, l =1, 2}
\phi_{k}  (z; \lambda ) d \rho _{kl} (\lambda) _{} \phi_{l}  (z^{\prime}; \lambda)^{}
=
\delta (z -z^{\prime})\:, 
\label{eq:completenessrelation0}
\end{eqnarray}
in the sense of (\ref{eq:weylstone}).  Eq. (\ref{eq:completenessrelation0}) being substituted into (\ref{eq:beforeusingcompleteness}), we are able to go  back to the defining equation (\ref{eq:greensfunction}) of the Green's function. It is obvious that   (\ref{eq:integralrepresentation}) also satisfies  (\ref{eq:greensfunction2}).

While it had been shown in \cite{weyl} that $ \rho_{kl}(\lambda_{2}) - \rho_{kl} (\lambda_{1} )$ is a positive semi-definite matrix for $\lambda_{2} >   \lambda_{1} $,  
Titchmarsh \cite{titchmarsh} and Kodaira \cite{kodaira1} went one step  further to derive the relations 
\begin{eqnarray}
\rho_{11}(\lambda_{2}) - \rho_{11}(\lambda_{1})
&=&
\lim_{\varepsilon \to +0} \frac{1}{\pi } \int _{\lambda_{1}}^{\lambda_{2} }
\Im \left (
\frac{1}{m_{1}(\lambda + i \varepsilon)  - m_{2}(\lambda + i \varepsilon) }
\right ) d\lambda  \: ,
\label{eq:titchmarskodaira1}
\\
\rho_{12}(\lambda_{2}) - \rho_{12}(\lambda_{1})
&=&
\rho_{21}(\lambda_{2}) - \rho_{21}(\lambda_{1})
\nonumber \\
&=&
\lim_{\varepsilon \to +0} \frac{1}{2 \pi } \int _{\lambda_{1}}^{\lambda_{2} }
\Im \left (
\frac{ m_{1}(\lambda + i \varepsilon) +
m_{2}(\lambda + i \varepsilon) 
}{m_{1}(\lambda + i \varepsilon)  - m_{2}(\lambda + i \varepsilon) }
\right ) d\lambda \:,
\label{eq:titchmarskodaira2} \\
\rho_{22}(\lambda_{2}) - \rho_{22}(\lambda_{1})
&=&\lim_{\varepsilon \to +0} \frac{1}{\pi } \int _{\lambda_{1}}^{\lambda_{2} }
\Im \left (
\frac{ m_{1}(\lambda + i \varepsilon)  \:  m_{2}(\lambda + i \varepsilon) }{m_{1}(\lambda + i \varepsilon)  - m_{2}(\lambda + i \varepsilon) }
\right ) d\lambda \:, 
\label{eq:titchmarskodaira3}
\end{eqnarray}
and thereby paved the way for determining the  spectral function  $\rho_{kl}(\lambda)$.  Here the symbol $\Im$ denotes the imaginary part and the  two quantities $m_{1}(\lambda + i\varepsilon )$ and 
$m_{2}(\lambda +  i\varepsilon )$ are called characteristic functions  in \cite{kodaira1}. The residue theorem in the complex function theory is made use of extensively in \cite{titchmarsh}, while  the spectral theorem for  the self-adjoint differential operator plays the key role in \cite{kodaira1}.   See Eqs.(3.1.5) - (3.1.8) in Ref.\cite{titchmarsh}  and  Eq.(1.15) in Ref.\cite{kodaira1} .

Weyl   classified the properties of the boundary points ($z=+\infty $ and $z=-\infty$  in the present case)  into two types,  ``limit circle type'' and   ``limit point type'' \cite{weyl}. If both of $z=+\infty$ and $z=-\infty$ are the limit point type, then the characteristic functions are given implicitly   in  \cite{titchmarsh}  and  explicitly  in \cite{kodaira1}  in the following  form
 \begin{eqnarray}
 m_{1}(\lambda + i\varepsilon )&=& - \lim_{a \to  - \infty}\frac{\phi_{1}(a; \lambda + i \varepsilon )}
 {\phi_{2} ( a ; \lambda + i \varepsilon )}
 \:,
 \hskip0.3cm 
  m_{2}(\lambda + i\varepsilon )= - \lim_{b \to  + \infty}\frac{\phi_{1}(b; \lambda + i \varepsilon )}
  {\phi_{2} ( b ; \lambda + i \varepsilon  )}\:.
  \label{eq:m1m2formulae}
 \end{eqnarray}
 See also Refs.  \cite{yoshida, yosida2} for an alternative and simpler derivation  of (\ref{eq:m1m2formulae}).  Procedures  of computing the characteristic functions  for the case of the limit circle type   have also been worked out, although a little involved. For more details, see \cite{kodaira1}.

 If   $M(z)^{2}$ is such that the Schr{\" o}dinger type equation (\ref{eq:eigenvalueproblem}) admits 
bound state solutions,  the spectral functions increase stepwise as $\lambda$ goes up through the corresponding binding energy level.  If there is no bound states, on the other hand,  the spectral functions are continuous and differentiable.  Then the Stieltjes integral  in (\ref{eq:integralrepresentation})   turns out to be the ordinary Riemann   integral over $\lambda$ with the following spectral weight, 
\begin{eqnarray}
d \rho_{11}(\lambda_{}) 
&=&
\lim_{\varepsilon \to +0} \frac{1}{\pi } 
\Im \left (
\frac{1}{m_{1}(\lambda + i \varepsilon)  - m_{2}(\lambda + i \varepsilon) }
\right ) d\lambda  \: ,
\label{eq:titchmarskodaira4}
\\
d \rho_{12}(\lambda_{}) 
&=&
d \rho_{21}(\lambda_{}) 
\nonumber \\
&=&
\lim_{\varepsilon \to +0} \frac{1}{2\pi } 
\Im \left (
\frac{ m_{1}(\lambda + i \varepsilon) + m_{2}(\lambda + i \varepsilon)  }{m_{1}(\lambda + i \varepsilon)  - m_{2}(\lambda + i \varepsilon) }
\right ) d\lambda \:,
\label{eq:titchmarskodaira5} 
\\
d \rho_{22}(\lambda_{}) 
&=&
\lim_{\varepsilon \to +0} \frac{1}{\pi } 
\Im \left (
\frac{ m_{1}(\lambda + i \varepsilon)  \:  m_{2}(\lambda + i \varepsilon) }{m_{1}(\lambda + i \varepsilon)  - m_{2}(\lambda + i \varepsilon) }
\right ) d\lambda \:. 
\label{eq:titchmarskodaira6}
\end{eqnarray}
The formulae   (\ref{eq:m1m2formulae}) -  (\ref{eq:titchmarskodaira6}) tell us  that we can evaluate the spectral density  $d\rho_{kl}(\lambda )/ d\lambda $,  once we know the asymptotic behavior at infinity  of the fundamental solutions defined by   (\ref{eq:initialconditions1})  and (\ref{eq:initialconditions}). 
This is the key observation in the spectral theory.

 \subsection{The constant mass case (no bubble wall)   }
 \label{subsectionzindependentcase}
 
 As the simplest exercise,  let us consider the  well-known case,  i.e.,  $M (z) ^{2}=M^{2}$ (constant), for which there is  no bubble wall. The fundamental solutions to (\ref{eq:eigenvalueproblem}) satisfying (\ref{eq:initialconditions1}) and (\ref{eq:initialconditions})  for $\lambda > M^{2}$ are simply given by 
 \begin{eqnarray}
 \phi_{1}(z; \lambda )={\rm cos}\left ( \sqrt{\lambda - M^{2}}\: z \right )\:,
 \hskip0.5cm
  \phi_{2}(z; \lambda )=\frac{1}{\sqrt{\lambda - M^{2}}\:}{\rm sin}\left ( \sqrt{\lambda - M^{2}}\: z \right )\:,
 \label{eq:constantmasseigenfunction}
 \end{eqnarray}
 and those  for $\lambda < M^{2}$  are obtained by analytic continuation of (\ref{eq:constantmasseigenfunction}) with respect to $\lambda$ or more explicitly are given by 
 \begin{eqnarray}
 \phi_{1}(z; \lambda )={\rm cosh}\left ( \sqrt{ M^{2} - \lambda }\: z \right )\:,
 \hskip0.5cm
  \phi_{2}(z; \lambda )=\frac{1}{\sqrt{ M^{2} - \lambda }\:}{\rm sinh}\left ( \sqrt{ M^{2} - \lambda }\: z \right )\:,
 \end{eqnarray}
 The characteristic functions are computed by using  (\ref{eq:m1m2formulae})  and we get 
 \begin{eqnarray}
 m_{1}(\lambda )=-i \sqrt{\lambda - M^{2}}\:, 
 \hskip0.5cm
 m_{2}(\lambda )= i \sqrt{\lambda - M^{2}}\:, 
\hskip1cm  {\rm for \:\:\:\:} \lambda > M^{2},
 \\
  m_{1}(\lambda )= \sqrt{ M^{2} - \lambda }\:, 
 \hskip0.5cm
 m_{2}(\lambda )= - \sqrt{ M^{2} - \lambda }\:,
\hskip1cm  {\rm for \:\:\:\:} \lambda <  M^{2}\:.
 \end{eqnarray}
 Therefore the   integral in (\ref{eq:integralrepresentation}) should be done by setting 
 \begin{eqnarray}
 d\rho_{11}(\lambda)
 &=& 
 \left \{
 \begin{tabular}{l}
 $\displaystyle{\frac{1}{2\pi}\frac{1}{\sqrt{\lambda - M^{2}}} \: d\lambda }\:,$ \hskip1cm for $\lambda > M^{2}\: ,$
 \\
\\
0 \hskip3.6cm for $ \lambda < M^{2}\:,$
\end{tabular}
 \right . 
  \label{eq:stieltjesmeasure1}
 \\
 d\rho_{12}(\lambda)&=& d\rho_{21}(\lambda)=0,
  \label{eq:stieltjesmeasure2}
   \\
 d\rho_{22}(\lambda)&=& 
 \left \{
 \begin{tabular}{l}
 $\displaystyle{ \frac{1}{2\pi} \sqrt{\lambda - M^{2}} \: d\lambda}   \:,$ \hskip1cm for $\lambda > M^{2}\:,$
 \\
 \\
 0 \hskip3.6cm for $\lambda < M^{2}\:.$
 \end{tabular}
 \right .
 \label{eq:stieltjesmeasure}
 \end{eqnarray}
 Note that the integration region  in (\ref{eq:integralrepresentation})  is automatically restricted to $\lambda >  M^{2}$, by virtue of  the above  formulae. 

 Combining the solutions   (\ref{eq:constantmasseigenfunction})  with    (\ref{eq:stieltjesmeasure1}),     (\ref{eq:stieltjesmeasure2})  and   (\ref{eq:stieltjesmeasure})  we get 
 \begin{eqnarray}
\int  \sum_{k, l  =1,2}\phi_{k}(z;\lambda)d\rho_{kl}(\lambda)\phi_{l}(z^{\prime};\lambda)
 &=&
 \frac{1}{2\pi} \int _{M^{2}}^{\infty} {\rm cos}\left \{
 \sqrt{\lambda - M^{2}}\: (z-z^{\prime})
 \right \} \frac{d\lambda}{\sqrt{\lambda -M^{2}}} 
 \nonumber \\
&=&
 \frac{1}{2\pi} \int _{0}^{\infty} \left \{
e^{ip_{z} (z-z^{\prime} )  }
+
e^{ - ip_{z} (z-z^{\prime} )  }
 \right \}    dp_{z}
 \nonumber \\
 &=&
 \delta ( z - z^{\:\prime})\:, 
 \end{eqnarray}
 where we set    $p_{z}=\sqrt{\lambda -M^{2}}$.  We can easily confirm in this example that our Green's function  (\ref{eq:integralrepresentation}) turns out to be   the familiar  Feynman's scalar field propagator, i.e., 
 \begin{eqnarray}
\Delta_{F} \left (  x-x^{\prime} ; M  \right ) 
&=&
 \int \frac{d^{4}p\: }{(2\pi)^{4}}
e^{ - ip \cdot (x-x^{\prime})}
\frac{i}{ p^{2} -M^{2} + i \varepsilon  }\;.
\label{}
\end{eqnarray}
The method of the general expansion method explained in Sec. \ref{subsec:wstc} is thus reduced for the constant mass case to the familiar  Fourier expansion. If the mass is position-dependent, however, the Fourier expansion is no more useful and the general expansion method  in Sec. \ref{subsec:wstc}  is an appropriate tool  for  diagrammatical  analyses.

\subsection{A solvable  model of the bubble wall }
\label{subsec:solvablemodel(II)}

\begin{figure}[htb]
\begin{center}
  \includegraphics[width=10cm]{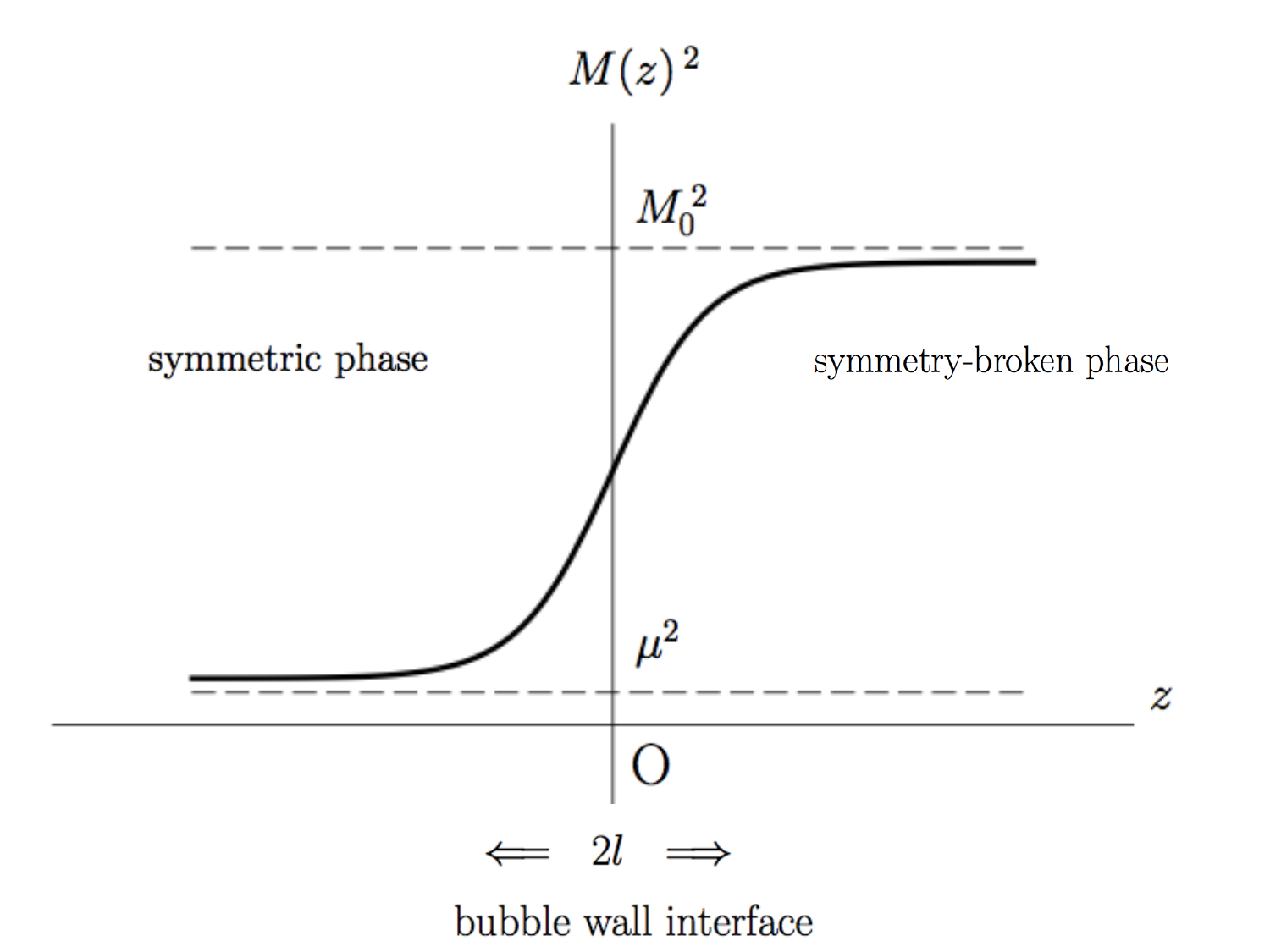}
\vskip0.3cm
\caption{ The gross shape of the  bubble wall given by (\ref{eq:bubblewall(II)}) with $u=0$.  }
\label{figure:wall}
\end{center}
\end{figure}

The second example  which is analytically solvable is given  by the following mass function
\begin{eqnarray}
M(z)^{2}&=&\mu^{2}(1-s)+ {M_{0}}^{2}\:s + u\:s(1-s)\: , 
\label{eq:bubblewall(II)}
\end{eqnarray}
where we have introduced a new variable 
\begin{eqnarray}
s=\frac{1}{2}\left \{  1+{\rm tanh}\left (  \frac{z}{2l}  \right )  \right\}
\label{eq:newvariables}
\end{eqnarray}
in place of $z$, thereby mapping  the whole $z$-space  ($- \infty < z < +\infty  $) into the finite region $0 < s < 1$.  This model has once been called attention to  in \cite{polyakov1} and was studied   thoroughly    in Refs. \cite{ayala,farrar}  in connection with a scattering problem in the presence of a domain wall.  The profile of the bubble wall is described by four parameters, $ \mu^{2}$, ${M_{0}}^{2}$, $u$,  and $l$.  The gross shape of the bubble wall is illustrated in Figure \ref{figure:wall} for $u=0$.   Note that for small values of   $u$, ${M(z)}^{2}$ is monotonically increasing and 
\begin{eqnarray}
\lim_{z \to +\infty} {M(z)}^{2} = {M_{0}}^{2}, 
\hskip1cm 
\lim_{z \to -\infty} {M(z)}^{2} = {\mu}^{2}\:.
\end{eqnarray}
The size of the bubble wall interface is determined by the length scale  $l$.  The  parameter $M_{0}$ is the mass observed in the symmetry-broken  phase and $\mu$ is expected to play the role of infrared regularization   and should be taken to be zero eventually. 

Let us seek for  the solutions to  (\ref{eq:eigenvalueproblem}) by setting 
\begin{eqnarray}
\phi_{k}(z; \lambda)
&=&s^{\kappa_{0} }(1-s)^{\kappa_{1}}\psi_{k}\left ( s; \lambda  \right )\:,   \hskip1cm ( k = 1,  2)\:.
\label{eq:seekforsolutions}
\end{eqnarray}
The two exponents $\kappa_{0}$ and $\kappa_{1}$ are determined by 
\begin{eqnarray}
{\kappa_{0}}^{2}
&=&
\left ( \mu^{2} -\lambda \right ) l^{2}, 
\hskip1cm 
{\kappa_{1}}^{2}
=
\left (  {M_{0}}^{2}- \lambda \right ) l^{2}\:,
\end{eqnarray}
and the choice of Riemann sheets when taking the square root will be discussed later.  Eq. (\ref{eq:eigenvalueproblem})    is now transformed into the following form
\begin{eqnarray}
0
&=&
\left \{ -\frac{d^{2}}{dz^{2}} + {M(z)}^{2} - \lambda   \right \}\phi_{k}(z; \lambda)
\nonumber \\
&=&
 - \frac{1}{l^{2}}  \: s^{\kappa_{0}+1}\: (1-s)^{\kappa_{1} + 1 } \bigg [  s(1-s)\frac{d^{2}}{ds^{2}}
+ \left \{  \gamma - \left (  \alpha + \beta + 1  \right )s \right \}   \frac{d}{ds} -\alpha \beta  \bigg ] 
\psi_{k}\left ( s; \lambda \right )\:, 
\nonumber \\
\label{eq:gausshypergeometric}
\end{eqnarray}
where we have defined $\alpha $,  $\beta $ and $\gamma $ as 
\begin{eqnarray}
\alpha
&=&
\kappa_{0}+ \kappa _{1}+ \frac{1}{2} +  \delta \:,
 \\
\beta&=&\kappa_{0}+ \kappa _{1}+ \frac{1}{2} - \delta \:,
 \\
\gamma &=&1+2\kappa_{0}\:, 
\\
\delta &=&\sqrt{\frac{1}{4}-ul^{2}} \:.
\end{eqnarray}
Note that $\kappa_{0}$ and $\kappa_{1}$ together with $\alpha $ , $\beta$ and $\gamma $ all  depend on $\lambda $.  

Eq. (\ref{eq:gausshypergeometric}) is the  Gauss's hypergeometric equation and there are two independent solutions around $s=0$, one of which is given by
\begin{eqnarray}
F(\alpha, \beta , \gamma ; s) = 1 + \frac{\alpha \beta }{\gamma } s +
\frac{\alpha (\alpha +1) \beta ( \beta +1)}{\gamma (\gamma +1)} \frac{s^{2}}{2!} + \cdots  \cdots .
 \label{eq:basissolutions0}
\end{eqnarray}
There are several ways of expressing  the second solution and here  we  take the following  one,
\begin{eqnarray}
 s^{1-\gamma}  F( 1 + \alpha - \gamma , 1 + \beta - \gamma , 2-\gamma ; s) \:.
  \label{eq:basissolutions}
\end{eqnarray}
Thus we express $\psi_{k}( s ; \lambda )$ as a linear combination of (\ref{eq:basissolutions0})  and  (\ref{eq:basissolutions}), namely, 
\begin{eqnarray}
\psi_{k}(s; \lambda)
&=&
c_{k1}(\lambda)\: F(\alpha , \beta , \gamma ; s) 
\nonumber \\
& & + c_{k2}(\lambda)\:s^{1-\gamma} 
F( 1 + \alpha - \gamma , 1 + \beta - \gamma , 2-\gamma ; s) \:,
\hskip0.5cm  (k=1, 2)\:.
\label{eq:linearcombination}
\end{eqnarray}

Since we are interested in the characteristic functions (\ref{eq:m1m2formulae}), we ought to examine the $ z \to - \infty$ and $z \to +\infty$ behavior of (\ref{eq:seekforsolutions})  or equivalently $s \to 0$ and $s \to 1$ behavior of (\ref{eq:linearcombination}). The first term of (\ref{eq:linearcombination}) in the   $s \to 0$ limit is simply determined  by the formula $F(\alpha , \beta , \gamma ; 0)=1$, while the behavior of the second term depends on the factor $s^{1-\gamma}$.  Here we require the following inequality for the real part ($\Re$)
 of the exponent
\begin{eqnarray}
\Re (1-\gamma )= -2 \Re ( \kappa_{0} ) > 0 \:,
\end{eqnarray}
under which  the first term in  (\ref{eq:linearcombination}) is more dominant than the second as $ s \to 0$.   This indicates that we should choose $\kappa_{0}$ as 
\begin{eqnarray}
\kappa_{0} = 
\left \{
\begin{tabular}{l}
$-\sqrt{\mu^{2} - \lambda }\: l$ \hskip1.6cm {\rm for} \hskip0.3cm $\mu^{2} > \lambda$\:,
\\
$i \sqrt{ \lambda + i \varepsilon - \mu^{2}}\: l$   \hskip1cm {\rm for} \hskip0.3cm $\lambda > \mu^{2} $\:,
\end{tabular}
\right .
\end{eqnarray} 
where we always take the $\varepsilon  \to + 0$ limit. 

Next let us turn to the $s \to 1$   limit of  (\ref{eq:linearcombination}). We know that 
$F(\alpha , \beta , \gamma ; 1)$ is finite and is given by the Gauss's formula
\begin{eqnarray}
F(\alpha , \beta , \gamma ; 1) = \frac{\Gamma (\gamma ) \Gamma ( \gamma - \alpha - \beta )}{\Gamma (\gamma - \alpha ) \Gamma (\gamma - \beta) }\;,  
\end{eqnarray} 
if  $\Re \gamma >0 $  and  the following inequality is satisfied
\begin{eqnarray}
\Re ( \gamma - \alpha - \beta ) =-2 \Re (\kappa_{1}) > 0\:.
\label{kappa1unequality}
\end{eqnarray}
Note that $F(1+\alpha - \gamma , 1+\beta -\gamma, 2-\gamma ; s)$  in the second term of 
(\ref{eq:linearcombination}) is also finite as $s \to 1$ 
if $\Re (2- \gamma ) > 0$ and  $\Re ( \gamma - \alpha - \beta ) > 0$.   Thus 
in order to meet the inequality  (\ref{kappa1unequality}), 
$\kappa_{1}$ has to be
\begin{eqnarray}
\kappa_{1} = 
\left \{
\begin{tabular}{l}
$-\sqrt{{M_{0}}^{2} - \lambda }\: l$ \hskip1.6cm {\rm for} \hskip0.3cm ${M_{0}}^{2} > \lambda$\:,
\\
$i \sqrt{ \lambda + i \varepsilon - {M_{0}}^{2}}\: l$   \hskip1cm {\rm for} \hskip0.3cm $\lambda > {M_{0}}^{2} $\:. 
\end{tabular}
\right .
\end{eqnarray} 
With these choices of the Riemann sheets,  the characteristic functions  (\ref{eq:m1m2formulae}) turn out to be 
\begin{eqnarray}
m_{1}(\lambda  )
&=&
 - \frac{c_{11}(\lambda )}{c_{21}(\lambda  )}\:,
\nonumber \\
m_{2}(\lambda  )
&=& - \frac{
c_{11}(\lambda   )\: F(\alpha , \beta , \gamma ; 1) 
 + c_{12}(\lambda   )\:
F( 1 + \alpha - \gamma , 1 + \beta - \gamma , 2-\gamma ; 1) 
}{
c_{21}(\lambda   )\: F(\alpha , \beta , \gamma ; 1) 
 + c_{22}(\lambda   )\:
F( 1 + \alpha - \gamma , 1 + \beta - \gamma , 2-\gamma ; 1) 
}\:.
\nonumber \\
\label{eq:characteristicfunctionsm1m2}
\end{eqnarray}

The coefficients $c_{kl}( \lambda )$  in (\ref{eq:linearcombination})  are determined by the conditions  (\ref{eq:initialconditions1})   and  (\ref{eq:initialconditions})  at $z=0$ (or equivalently at $s=1/2$), and we list up the results:
\begin{eqnarray}
c_{11}(\lambda ) &=&\frac{2^{ \kappa_{0}+\kappa_{1}  + \gamma -1}}{W}
\bigg [
- 2(\kappa_{0} + \kappa_{1}) F(\alpha -\gamma +1, \beta - \gamma +1, 2-\gamma ; 1/2 )
\nonumber \\
& & 
\hskip3cm
+F^{\;\prime} (\alpha - \gamma +1, \beta - \gamma +1, 2-\gamma ;  1/2 )
\bigg ]  \:, 
\\
c_{12}(\lambda ) &=& - \frac{2^{\kappa_{0}+\kappa_{1}}}{W}\bigg [
2(\kappa_{0} - \kappa_{1}) F(\alpha , \beta , \gamma ; 1/2  )
+
F^{\:\prime } (\alpha , \beta , \gamma ; 1/2  )
\bigg ]\:,
\\
c_{21}(\lambda ) &=&
-\frac{2^{\kappa_{0} + \kappa _{1} + \gamma +1 }}{W}
F(\alpha - \gamma +1, \beta - \gamma +1, 2-\gamma ; 1/2  ) \:l\:,
\\
c_{22} ( \lambda ) &=&
\frac{2^{\kappa_{0} + \kappa _{1}  + 2 }}{W}
F(\alpha , \beta ,   \gamma ; 1/2   ) \:l\:.
\end{eqnarray}
Here $W$ in the denominator is given  by 
\begin{eqnarray}
W&=&
2^{\gamma -1}
\bigg [
F ( \alpha , \beta , \gamma ; 1/2 ) 
F^{\:\prime} ( \alpha -\gamma +1, \beta -\gamma +1, 2-\gamma ; 1/2  )
\nonumber \\
& &
- F ( \alpha - \gamma +1 , \beta - \gamma +1 , 2- \gamma ; 1/2 )
F^{\:\prime} ( \alpha , \beta , \gamma ; 1/2  )
\nonumber \\
& &
+2(1-\gamma )
F ( \alpha , \beta , \gamma ; 1/2 ) F ( \alpha - \gamma +1 , \beta - \gamma +1 , 2- \gamma ; 
 1/2  )
\bigg ]\:.
\end{eqnarray}

\begin{figure}[h]
\centering
  \includegraphics[width=12cm]{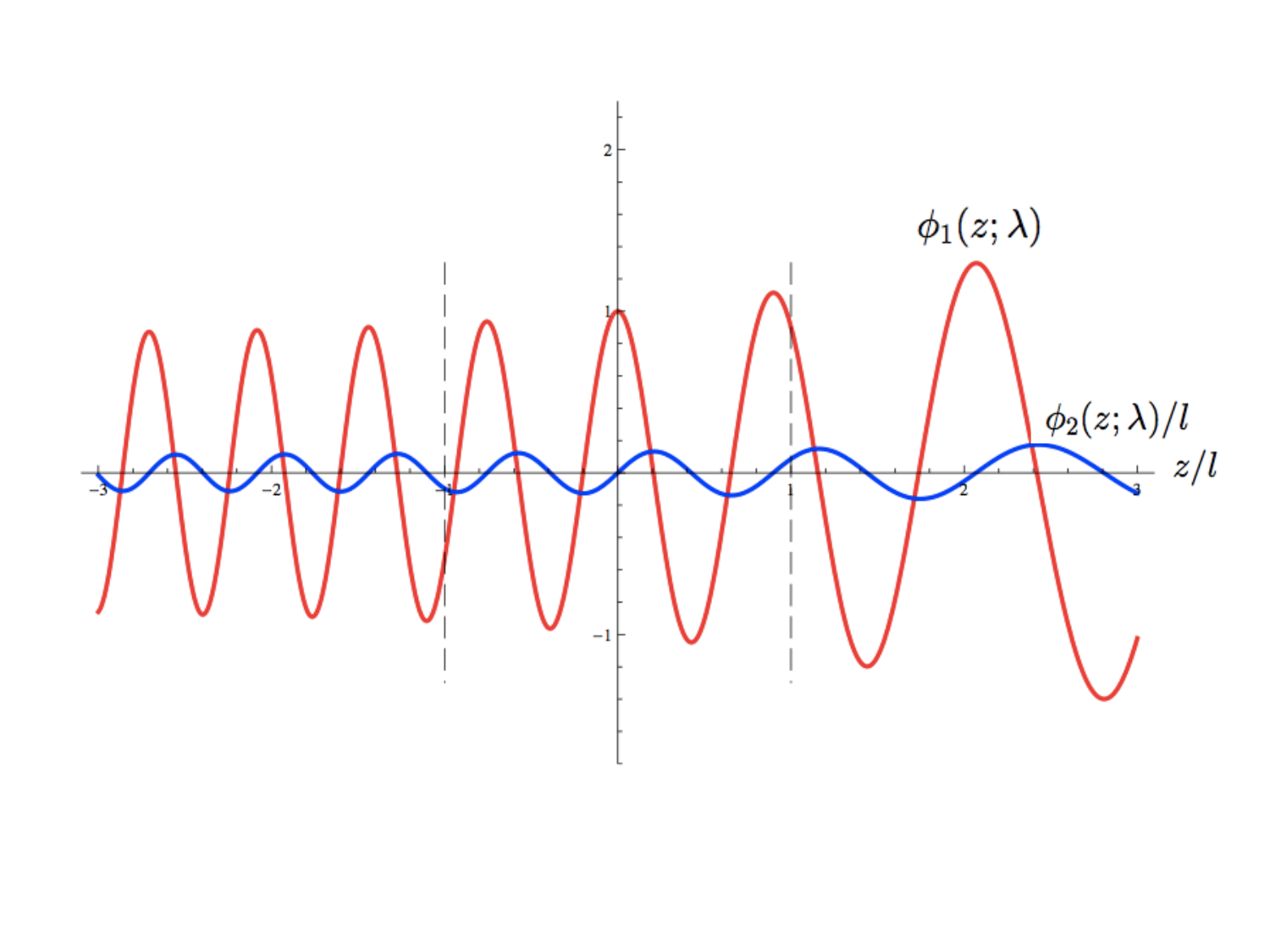}
\caption{
The typical behavior of the fundamental solutions $\phi_{1}(z;\lambda)$ (red curve) and $\phi_{2}(z;  \lambda )/l$ (blue curve) in the model  (\ref{eq:bubblewall(II)}) for $\lambda > {M_{0}}^{2} > \mu^{2}$.  Note that both   $\phi_{1}(z;\lambda)$   and    $\phi_{2}(z;  \lambda )/l$   are dimensionless.     We take   $1/l$   as the unit of the mass scale, and   other  parameters are chosen   as follows:  $M_{0}=10/l$, $\mu = 10^{-4}/l$,  $\lambda =110/l^{2}$  and $u$=0..  The region $-1 < z/l <  +1 $ sandwiched by the two dashed lines is the bubble wall interface.
 }
\label{figure:wstkfigure}
\end{figure}

With these coefficients,  we illustrate the behavior of $\phi_{1}(z;\lambda)$ and $\phi_{2}(z;  \lambda )$
in Figure \ref{figure:wstkfigure} for sample values of ${M_{0}}^{}$, $\mu^{}$, $\lambda$ and $u$ in the mass unit $1/l$.
Figure \ref{figure:wstkfigure}  exhibits oscillatory behaviors like  sinusoidal waves.  The hypergeometric function shows similar behavior in the finite $z$ region, but at infinity ( $z \to \pm \infty$,   i.e., $s \approx 0$ and $s \approx 1$ ), the prefactor $s^{\kappa_{0}} (1-s)^{\kappa_{1}} $ in Eq. (\ref{eq:seekforsolutions})  and $s^{1-\gamma} = s^{ - 2\kappa_{0}}$ in the second term in (\ref{eq:linearcombination})  are also responsible for such behaviors.  Note that both $\kappa_{0}$ and $\kappa_{1}$ are pure imaginary for our choice of the parameters in  Figure \ref{figure:wstkfigure}. For large $\vert z \vert $,   we have 
 \begin{eqnarray}
 (1- s) \sim e^{-z/l}\:,  \hskip1cm  (1- s)^{\kappa_{1}} \sim e^{-\kappa_{1}z/l} \hskip1cm &{\rm for}& \hskip0.5cm  z\to +\infty \:,
 \nonumber 
 \\
 s \sim e^{z/l}\:, \hskip2.3cm  s^{\kappa_{0}} \sim e^{\kappa_{0}z/l} \hskip1cm &{\rm for}& \hskip0.5cm  z\to -\infty  \:,
  \end{eqnarray}
  and therefore the large negative  $z$ behavior of $\phi_{k}(z; \lambda) $ is given by a linear combination of ${\rm cos}\:(\vert \kappa_{0} \vert z/l) $ and  ${\rm sin}\:(\vert \kappa_{0} \vert z/l) $.  Likewise  the large positive $z$ behavior is a  combination of ${\rm cos}\:(\vert \kappa_{1} \vert z/l) $ and  ${\rm sin}\:(\vert \kappa_{1} \vert z/l) $.
  
 The region sandwiched by the two dashed lines in Figure  \ref{figure:wstkfigure} ($-1 < z/l < +1$) is the bubble wall interface  and we can  see obviously that the wave length in the  $z \to + \infty $ region is longer than in the  $z \to -\infty $ region.  This means that the $z$-component of momentum carried by the propagating wave   is not conserved and some amount of $z$-momentum is transferred to the bubble wall when moving toward the $z \to +\infty$ region   through the interface.

With the explicit formulae   (\ref{eq:characteristicfunctionsm1m2}),  we can compute the $2 \times 2 $ spectral function matrix  $\rho_{kl}(\lambda )$   by using the formulae  (\ref{eq:titchmarskodaira1}), (\ref{eq:titchmarskodaira2})  and (\ref{eq:titchmarskodaira3}).  Without entering detailed calculations, however,  one thing  can be seen  clearly.   We  notice that, for ${M_{0}}^{2} > \mu^{2} > \lambda $, both $\kappa_{0}$ and $\kappa_{1}$ are real numbers  and that  $\alpha$, $\beta $ and $\gamma $ are all real numbers accordingly.   This tells us that there does not exist any source giving rise to  imaginary part on the right hand side of   (\ref{eq:titchmarskodaira1}), (\ref{eq:titchmarskodaira2})  and (\ref{eq:titchmarskodaira3}).  Therefore we can safely say that the  integration $d\rho_{kl}(\lambda )$  in 
(\ref{eq:integralrepresentation}) is restricted in the region $\lambda > \mu^{2}$. 

\begin{figure}[hbt]
\centering
 \includegraphics[width=10cm]{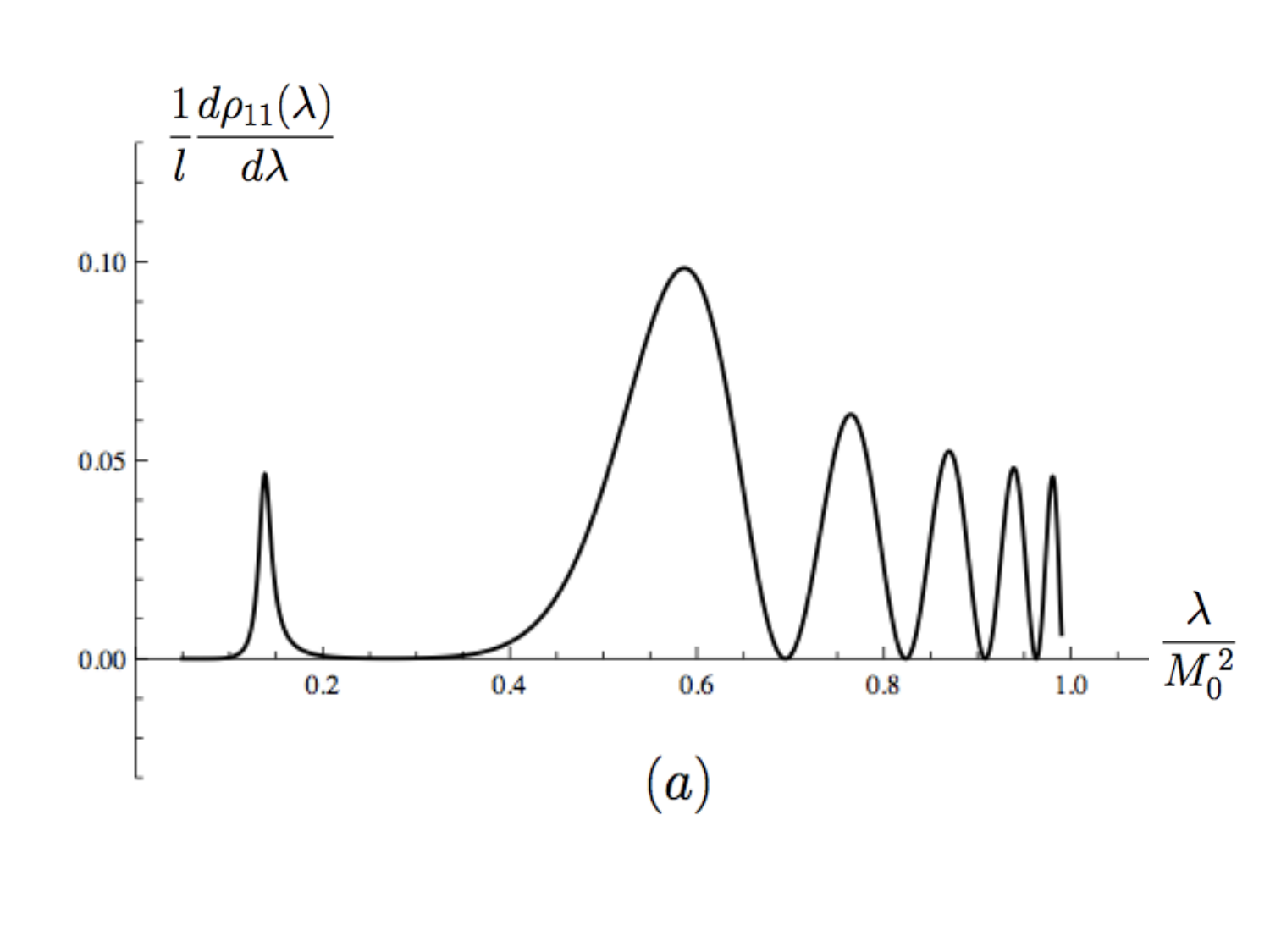}
 \includegraphics[width=10cm]{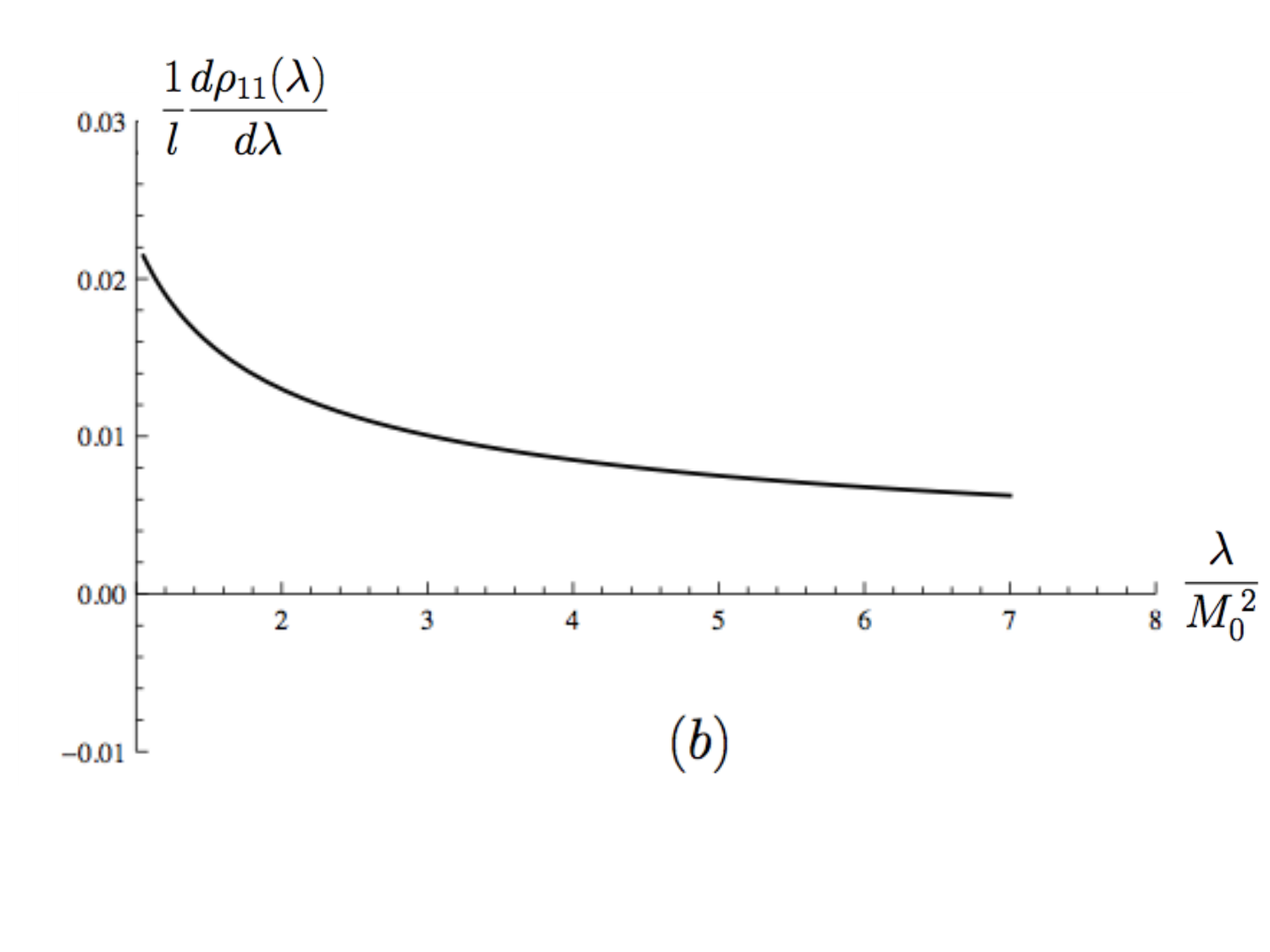}
\caption{
The spectral function $\displaystyle{\frac{1}{l}\frac{d\rho_{11}(\lambda )}{d\lambda}}$ 
for the region (a) $0.05 <  \lambda / M_{0}^{\:2}  < 0.99$ and (b) $1.05 < \lambda/M_{0}^{\:2} < 7.0$
By taking $1/l$ as the mass unit,  other  parameters are chosen   as   $M_{0}=10/l$,  $\mu = 10^{-4}/l$ and $u=0$. All of the numbers on the vertical and horizontal axes are dimensionless. 
 }
\label{figure:spectralfunction}
\end{figure}

The characteristic functions in the regions $M_{0}^{2} > \lambda > \mu^{2} $ and  $\lambda > M_{0}^{2} $ can be studied only  numerically and we illustrate the behavior of $d\rho_{11}(\lambda )/d\lambda $ in Figure \ref{figure:spectralfunction}. Looking at the behavior shown in Figure \ref{figure:spectralfunction} (a) for  the region $M_{0}^{2} > \lambda > \mu^{2} $, we immediately notice  the existence of  several ups and downs in sharp contrast with the smooth behavior in the region $\lambda > M_{0}^{2} $ in  Figure \ref{figure:spectralfunction} (b).  The  contrast of similar kind can be seen in $d\rho_{12}(\lambda )/d\lambda     =d\rho_{21}(\lambda )/d\lambda $   and $d\rho_{22}(\lambda )/d\lambda $.  Although   origins of this contrast have to be scrutinized further, this could be a warning message that we have to be careful of the soft  (i.e., small $\lambda $ )   region   integration in the loop diagrams. Large single and double logarithmic   effects are closely connected  with integration over  soft and/or collinear regions in phase space, and such a  behavior as in   Figure \ref{figure:spectralfunction} (a)   could  have  important influences in  analogous calculations in the presence of a bubble wall. 

\section{The Dirac Fermion's two-point Green's function}
\label{sec:fermiongreenfunction}
\subsection{Preliminaries}

Let us now move to the construction of the Fermion's propagator, whose mass is assumed to be $z$-dependent, i.e.,  $m(z)$. We are going to solve the following equation 
\begin{eqnarray}
& & 
\left \{ i \gamma ^{\mu} {\overrightarrow \partial }_{\mu}  - m(z) \right \}
{\cal F} \left ( t-t^{\: \prime}, x-x^{\: \prime}, y-y^{\: \prime}, z, z^{\: \prime} ; m(*) \right ) 
\nonumber \\
& & \hskip5cm
= i \: \delta(t-t^{\prime})\delta(x-x^{\: \prime}) \delta(y-y^{\: \prime})\delta(z-z^{\: \prime})\:,
\label{eq:fermionspropagator}
\\
& & 
{\cal F} \left ( t-t^{\: \prime}, x-x^{\: \prime}, y-y^{\: \prime}, z, z^{\: \prime} ; m(*) \right ) 
\left \{ i \gamma ^{\mu} {\overleftarrow \partial }^{\: \prime}_{\mu}  + m(z^{\:\prime}) \right \}
\nonumber \\
& & \hskip5cm
= - i \: \delta(t-t^{\prime})\delta(x-x^{\: \prime}) \delta(y-y^{\: \prime})\delta(z-z^{\: \prime})\:.
\label{eq:fermionspropagator2}
\end{eqnarray}
The derivatives in (\ref{eq:fermionspropagator}) and (\ref{eq:fermionspropagator2}) are given as usual by
\begin{eqnarray}
\partial_{\mu} = \left (
\frac{\partial }{\partial t}, \: \frac{\partial }{\partial x}, \: \frac{\partial }{\partial y}, \: \frac{\partial }{\partial z}
\right )\:, 
\hskip1cm 
\partial_{\mu}^{\:\prime} = \left (
\frac{\partial }{\partial t^{\:\prime}} , \: \frac{\partial }{\partial x^{\:\prime} }, \: \frac{\partial }{\partial y^{\:\prime} }, \: \frac{\partial }{\partial z^{\:\prime}}
\right )\:
\end{eqnarray}
and are  acting on the right ($\overrightarrow \partial_{\mu} $) and left ($\overleftarrow \partial_{\mu} $) quantities,   respectively.  As in the Bosonic case,  our notation $m(*)$ instead of $m(z)$ or $m(z^{\:\prime})$ in (\ref{eq:fermionspropagator}) and  (\ref{eq:fermionspropagator2})  is  to remind us that the two-point function  ${\cal F}$ depends  on the mass function only implicitly.    Note that  ${\cal F}$ carries two spinor indices that are, however,  suppressed in  (\ref{eq:fermionspropagator}) and (\ref{eq:fermionspropagator2}). Our propagator is necessarily a  generalization of the well-known  one 
\begin{eqnarray}
S_{F}(x - x^{\:\prime} )=i  \int \frac{d^{4}p }{(2\pi)^{4}}\: e^{-ip\cdot ( x - x^{\: \prime})} \: \frac{\gamma ^{\mu} p_{\mu} + m}{p^{2}-m^{2} + i \varepsilon }\:, 
\label{eq:constantmassfermion}
\end{eqnarray}
and is supposed to replace (\ref{eq:constantmassfermion}) in the presence of the electroweak bubble wall.

\subsection{The canonical form  }

The ways of handling the dynamics along the $z$-axis will differ from those in the $x$- and $y$-axes and  the most convenient  choice of the gamma-matrices that we found  is  rather unconventional, i.e., 
\begin{eqnarray}
& & \gamma^{0}=\left (
\begin{tabular}{cc}
${\bf 0}$&${\bf 1}$
\\
${\bf 1}$&${\bf 0}$
\end{tabular}
\right ), 
\hskip1.cm
\gamma^{1}=\left (
\begin{tabular}{cc}
$ {\bf 0}$&$\sigma^{2}$
\\
$-\sigma^{2}$ & ${\bf 0}$
\end{tabular}
\right )\:,
\nonumber \\
& & 
\gamma^{2}=\left (
\begin{tabular}{cc}
$ {\bf 0}$&$ - \sigma^{1}$
\\
$\sigma^{1}$ & ${\bf 0}$
\end{tabular}
\right )\:,
\hskip0.3cm
\gamma^{3}= -i \left (
\begin{tabular}{cc}
$ {\bf 1}$&${\bf 0}$
\\
${\bf 0 }$ & ${\bf -1}$
\end{tabular}
\right )\:.
\label{eq:gammamatrix}
\end{eqnarray}
Here ${\bf 1} $ and ${\bf 0}$ are the $2 \times 2$ unit matrix and zero-matrix, respectively.  The Pauli matrices are denoted by $\sigma^{i}  \:\:(i=1,2,3)$. This set of gamma-matrices is obtained by applying a unitary transformation to the Dirac's representation and is characteristic in  that the third Pauli matrix $\sigma^{3}$  does not appear in (\ref{eq:gammamatrix}). 

In constructing the Bosonic Green's function in Sec.\ref{sec:green}, we learned that  the spectral problem  of the second-order differential operator and in particular the determination of $d\rho_{kl}(\lambda )/d\lambda$  was our central concern.  We have to think about similar ways of formulation that is suitable for the first-order Dirac operator.   Fortunately, spectral problems for a first-order differential equation for a pair of functions   have been studied in mathematical literatures for the case of  half finite interval,  i.e.,  $z \in (0, +\infty ) $ in Refs.\cite{conte,roos,titchmarsh2} and for infinite open interval,  i.e., $z \in (-\infty, +\infty)$ in Ref. \cite{titchmarsh2}.   Self-contained monographs on this subject are also available \cite{levitan, levitan2}. Bearing these mathematical knowledge in our mind,  let us  disentangle  the Dirac equation 
\begin{eqnarray}
\left \{
i \gamma \cdot \partial  -m(z) 
\right \} \psi (z) \: e^{-iEt + i \vec{p}_{\perp} \cdot \vec{x}_{\perp}}
=
 \left \{
\gamma \cdot p -m(z) 
\right \} \psi  (z) \: e^{-iEt + i \vec{p}_{\perp} \cdot \vec{x}_{\perp}}
=  0    \:\:,
\nonumber \\
p^{\mu} = ( E, \: p_{x}, \: p_{y} , \: -i\frac{\partial }{\partial z}  )\:,  
\label{eq:eigenvalueproblemoriginal}
\end{eqnarray}
in a simpler form.  Here   $E$ and  $\vec{p}_{\perp}=(p_{x}, p_{y}, 0)$ are  just numbers, but 
the $z$-component of the momentum operator    gives rise 
to a non-trivial differential equation for $ \psi(z)$ due to the $z$-dependent mass $m(z)$.  
We now put  (\ref{eq:eigenvalueproblemoriginal}) into a form that is suitable  to    
the analyses in Ref. \cite{titchmarsh2}, 

By decomposing $\psi (z) $ into upper and lower  two-component spinors, $\xi_{1}$ and $\xi_{2}$, 
\begin{eqnarray}
\psi (z) =\left (
\begin{tabular}{c}
$\xi_{1}$
\\  
$\xi_{2}$
\end{tabular}
\right ),
\end{eqnarray}
the Dirac equation (\ref{eq:eigenvalueproblemoriginal})  now becomes
\begin{eqnarray}
 \left (  E + \sigma^{2} p_{x} -  \sigma^{1}p_{y}   
 \right )   \xi_{1}
+
\left \{ - \frac{d}{dz}  - m(z) \right \} \xi_{2} 
=0\:,
 \label{eq:reduceddirac1}
\\
\left \{  \frac{d}{dz}   - m(z) \right \} \xi_{1}
+
 \left (  E  -  \sigma^{2}p_{x} + \sigma^{1}p_{y}  \right )   \xi_{2}
 =0  \:.
 \label{eq:reduceddirac2}
\end{eqnarray}
Thanks  to the use of the gamma-matrix representation as in (\ref{eq:gammamatrix}), we find that the Pauli matrix appears in (\ref{eq:reduceddirac1}) and (\ref{eq:reduceddirac2}) only  in the combination of 
$ ( \sigma^{2}p_{x} - \sigma^{1} p_{y} ) $. This suggest that (\ref{eq:reduceddirac1}) and (\ref{eq:reduceddirac2}) will become much simpler if we use the eigenstates of $ ( \sigma^{2}p_{x} - \sigma^{1} p_{y} ) $ as the basis of our calculation.The eigenstates of $ ( \sigma^{2}p_{x} - \sigma^{1} p_{y} ) $  are given by 
\begin{eqnarray}
\eta^{(\pm)}=\frac{1}{\sqrt{2} \:  \vert \vec{p}_{\perp} \vert }\left (
\begin{tabular}{c}
$ \mp ( i\:p_{x}+p_{y})  $
\\
\\
$\vert \vec{p}_{\perp} \vert $
\end{tabular}
\right )\:, 
\hskip1cm
\vert {\vec p}_{\perp} \vert = \sqrt{p_{x}^{2} + p_{y}^{2}}\:,
\end{eqnarray}
which are shown  to  satisfy
\begin{eqnarray}
\frac{1}{\vert \vec{p}_{\perp} \vert }
\left (\sigma^{2}p_{x} - \sigma^{1} p_{y} \right)\eta^{(\pm)}
=
\pm \eta^{(\pm)}
\end{eqnarray}
together with 
\begin{eqnarray}
& & 
\eta^{(+) \dag } \eta^{(+)}=
\eta^{(-) \dag } \eta^{(-)}=1, \hskip0.5cm
\eta^{(+) \dag } \eta^{(-)}= \eta^{(-) \dag } \eta^{(+)}=0  \: ,
\\
& & \eta^{(+)}\eta^{(+) \dag}+ \eta^{(-)}\eta^{(-) \dag}={\bf 1}\:,
\label{eq:etaformula1}
\\
& & \eta^{(+)}\eta^{(+) \dag} -  \eta^{(-)}\eta^{(-) \dag}
=
\frac{1}{\vert \vec{p}_{\perp} \vert }
\left (\sigma^{2}p_{x} - \sigma^{1} p_{y} \right)\:.
\label{eq:etaformula2}
\end{eqnarray}

We now eliminate the matrix $ ( \sigma^{2}p_{x} - \sigma^{1}p_{y} ) $ 
from (\ref{eq:reduceddirac1}) and (\ref{eq:reduceddirac2})  
by setting $\xi_{1}$ and $\xi_{2}$ in the following way, 
\begin{eqnarray}
\xi_{1} =  f^{(\pm )} (z; E ) \: \eta^{(\pm)}\:,
\hskip2cm
\xi_{2} = g^{(\pm )} (z; E ) \: \eta^{(\pm)}\:.
\end{eqnarray}
Here our notations,  $f^{( \pm ) } (z; E )$ and $g^{ \pm ) } (z; E )$, are to keep in our mind that   these two functions constitute the solution  belonging to the energy  $E$.    They are determined by the coupled differential equation 
\begin{eqnarray}
 \left (  E \pm  \vert  \vec{p}_{\perp} \vert   \right )   f^{(\pm ) } (z; E ) 
+
\left \{ - \frac{d}{dz}  - m(z) \right \} g^{(\pm )} (z; E ) 
=0\:,
 \label{eq:reduceddirac3}
\\
\left \{  \frac{d}{dz}   - m(z) \right \} f^{(\pm )} (z; E ) 
+
 \left (  E  \mp  \vert  \vec{p}_{\perp} \vert       \right )   g^{(\pm )} (z; E ) 
 =0 \:.
 \label{eq:reduceddirac4}
\end{eqnarray}
This may be rewritten in the matrix form
\begin{eqnarray}
\left (
\begin{tabular}{cc}
$0$ &  $1$
\\
$-1$ &  $0$
\end{tabular}
\right )\frac{d}{dz}
\left (
\begin{tabular}{c}
$f^{(\pm)}(z; E )$
\\
$g^{(\pm)}(z; E )$
\end{tabular}
\right )+
\left (
\begin{tabular}{cc}
$\mp \vert \vec{p}_{\perp} \vert $ & $m(z)$
\\
$m(z)$ & $\pm \vert \vec{p}_{\perp} \vert $
\end{tabular}
\right )
\left (
\begin{tabular}{c}
$f^{(\pm)}(z; E )$
\\
$g^{(\pm)}(z; E )$
\end{tabular}
\right )
\nonumber \\
=E
\left (
\begin{tabular}{c}
$f^{(\pm)}(z; E )$
\\
$g^{(\pm)}(z; E )$
\end{tabular}
\right )\:.
\label{eq:standardform1}
\end{eqnarray}

Now we  refine Eq. (\ref{eq:standardform1}) further by applying an orthogonal transformation 
\begin{eqnarray}
\left (
\begin{tabular}{c}
$f^{(\pm)}(z; E )$
\\
$g^{(\pm)}(z; E )$
\end{tabular}
\right )
= 
\left (
\begin{tabular}{cc}
${\rm cos}\: \theta^{(\pm )}(z)$ & $ - {\rm sin}\:\theta^{ (\pm ) }(z) $
\\
$ {\rm sin}\:\theta^{ ( \pm )} (z) $ & ${\rm cos}\: \theta^{ ( \pm ) } (z) $
\end{tabular}
\right ) 
\left (
\begin{tabular}{c}
$F^{(\pm)}(z; E )$
\\
$G^{(\pm)}(z; E )$
\end{tabular}
\right )\:.
\label{eq:orthogonalrotation}
\end{eqnarray}
Here we set 
\begin{eqnarray}
{\rm tan}\: 2 \theta^{ ( \pm ) } (z)  = \mp \frac{m(z)}{\vert \vec{p}_{\perp} \vert }
\end{eqnarray}
to find Eq. (\ref{eq:standardform1}) transformed into 
\begin{eqnarray}
\left (
\begin{tabular}{cc}
$0$ &  $1$
\\
$-1$ &  $0$
\end{tabular}
\right )\frac{d}{dz}
\left (
\begin{tabular}{c}
$F^{(\pm)}(z; E )$
\\
$G^{(\pm)}(z; E )$
\end{tabular}
\right )+
\left (
\begin{tabular}{cc}
$ \mp Q(z)$ &   $0$
\\
$0$ & $\pm Q(z) $
\end{tabular}
\right )
\left (
\begin{tabular}{c}
$F^{(\pm)}(z; E )$
\\
$G^{(\pm)}(z; E )$
\end{tabular}
\right )
\nonumber \\
= E
\left (
\begin{tabular}{c}
$F^{(\pm)}(z; E )$
\\
$G^{(\pm)}(z; E  )$
\end{tabular}
\right )\:,
\label{eq:standardform2}
\end{eqnarray}
with
\begin{eqnarray}
Q(z)=\sqrt{m(z)^{2} + \vert \vec{p}_{\perp} \vert^{2} }  
+ \frac{m^{\prime}(z) \vert \vec{p}_{\perp} \vert }{2\{   m(z)^{2} + \vert \vec{p}_{\perp} \vert ^{2} \} }\:.
\label{eq:defofq(z)}
\end{eqnarray}
This simplified form of the Dirac equation  is called ``canonical''   in Ref. \cite{levitan},   where the spectra  of the energy  $E$ were fully analyzed for   general ``potential''  $Q(z)$.  This canonical form     (\ref{eq:standardform2}) is useful to look into the symmetry property  of the $E$ spectra. In fact we can rewrite (\ref{eq:standardform2}) as 
\begin{eqnarray}
\left (
\begin{tabular}{cc}
$0$ &  $1$
\\
$-1$ &  $0$
\end{tabular}
\right )\frac{d}{dz}
\left (
\begin{tabular}{c}
$G^{(\pm)}(z; E )$
\\
$F^{(\pm)}(z; E )$
\end{tabular}
\right )+
\left (
\begin{tabular}{cc}
$ \mp Q(z)$ &   $0$
\\
$0$ & $\pm Q(z) $
\end{tabular}
\right )
\left (
\begin{tabular}{c}
$G^{(\pm)}(z; E )$
\\
$F^{(\pm)}(z; E )$
\end{tabular}
\right )
\nonumber \\
= - E
\left (
\begin{tabular}{c}
$G^{(\pm)}(z; \lambda )$
\\
$F^{(\pm)}(z; \lambda )$
\end{tabular}
\right )\:,
\label{eq:standardform3}
\end{eqnarray}
and we note the sign change,  $E \to - E$,  on the right hand side of (\ref{eq:standardform3}).  
In other words,  the coupled equation   (\ref{eq:standardform3}) for a pair of functions 
$F^{(\pm)}(z; \lambda )$ and  $G^{(\pm)}(z; \lambda )$  is form-invariant   under the following replacement
 \begin{eqnarray}
 E \to -E \;, 
 \hskip1cm
  \left (
\begin{tabular}{c}
$F^{(\pm)}(z; E )$
\\
$G^{(\pm)}(z; E )$
\end{tabular}
\right )
\to 
 \left (
\begin{tabular}{c}
$G^{(\pm)}(z;  E )$
\\
$F^{(\pm)}(z;  E )$
\end{tabular}
\right )\:.
 \end{eqnarray}
 This is the symmetry between positive and negative energy spectra in the usual sense.

 \subsection{The transverse momentum dependence of the energy $E$}
 
The energy $E$ that appeared in (\ref{eq:standardform2}) is inter-connected with  the magnitude of the 
transverse momentum $\vert \vec{p}_{\perp} \vert $. This connection can be made more explicit in the following way.   Let us now deduce second order differential equations for $f^{(\pm)}(z; E)$ and $g^{(\pm)}(z; E)$  from  (\ref{eq:reduceddirac3}) and (\ref{eq:reduceddirac4}),   which turn out to be the following one-dimensional Schr{\" o}dinger type  equations,  
\begin{eqnarray}
\left \{  -\frac{d^{2}}{dz^{2}} + U_{f}(z)  \right \}  f^{(\pm)}(z; E) &=& \nu_{f} f^{(\pm)}(z; E)\:,
\\
\left \{  -\frac{d^{2}}{dz^{2}} + U_{g}(z)  \right \}  g^{(\pm)}(z; E) &=& \nu_{g} g^{(\pm)}(z; E)\:.
\end{eqnarray}
Here  the ``potential''  terms are given by
\begin{eqnarray}
U_{f}(z) =m(z)^{2} + \frac{d m(z)}{dz}\:, 
\hskip1cm
U_{g}(z) =m(z)^{2} - \frac{d m(z)}{dz}\:, 
\end{eqnarray}
respectively and the ``energy''  $\nu_{f}$ and $\nu_{g}$,  of these potential problems are connected 
with $E$ by a common relation  
\begin{eqnarray}
\nu_{f}=E^{2} -\vert \vec{p}_{\perp} \vert^{2}\:,
\hskip1cm
\nu_{g}=E^{2} -\vert \vec{p}_{\perp} \vert^{2}\:.
\end{eqnarray}
In general  the allowed region of the spectra of $ \{ \nu_{f} \} $ and $\{ \nu_{g} \} $ do not 
necessarily coincide  with each other since the potentials are different.  Only when they share common 
region of the spectra i.e., $\nu_{f}=\nu_{g}=\nu$,  we are able to obtain 
the energy $E$ of  the coupled equations  (\ref{eq:reduceddirac3}) and (\ref{eq:reduceddirac4}) 
by the formula
\begin{eqnarray}
E = \pm \sqrt{\nu + \vert \vec{p}_{\perp} \vert ^{2}}\:.
\label{eq:restrictedform}
\end{eqnarray}
Here it should be stressed that the spectra of $ \{ \nu_{f} \} $ and $\{ \nu_{g} \} $ 
are independent of   $\vert \vec{p}_{\perp} \vert $ since the potentials,  
$U_{f}(z)$  and $U_{g}(z)$,  do not contain   $\vert \vec{p}_{\perp} \vert $. 
The energy  $E$ in   (\ref{eq:reduceddirac3}),  (\ref{eq:reduceddirac4}) and   
    (\ref{eq:standardform2}) depend on 
 $\vert \vec{p}_{\perp} \vert $ only through  the restricted form of (\ref{eq:restrictedform}) 
irrespectively  of the details of the form of $m(z)$.  This observation could be useful when we look at infrared singularities associated with loop diagrams, since singularities always come  from  limited corners in the phase space, i.e.,   $\vert \vec{p}_{\perp} \vert^{2} \approx 0$

 \subsection{The general expansion method and completeness}
  
 Let us now discuss an expansion of a pair of arbitrary real-valued functions  $ ( A(z) \:\:  B(z) )$  in terms of the set of solutions  $ ( F_{}^{(\pm)}(z; E ) \:\: G_{}^{(\pm)}(z; E ))$.   We assume that $A(z)$ and $B(z)$ are both {\it square-integrrable} in $(-\infty , +\infty )$.     We introduce two types of solutions to Eq. (\ref{eq:standardform2}) 
 $(  F_{1}^{(\pm)}(z; E ) \:\:\:   G_{1}^{(\pm)}(z; E )  )$ and 
  $(  F_{2}^{(\pm)}(z; E ) \:\:\:   G_{2}^{(\pm)}(z; E )  )$, 
 satisfying the conditions at $z=0$, 
 \begin{eqnarray}
& &  F_{1}^{(\pm)}(0; E )=1, 
\hskip1cm
 G_{1}^{(\pm)}(0; E) =0\:,
 \label{eq:fermionicbasicfunctions1}
\end{eqnarray}
and
\begin{eqnarray}
& &  F_{2}^{(\pm)}(0; E )=0, 
\hskip1cm
 G_{2}^{(\pm)}(0; E) =1\:,
  \label{eq:fermionicbasicfunctions2}
  \end{eqnarray}
analogously to Eqs. (\ref{eq:initialconditions1}) and  (\ref{eq:initialconditions}) in the Bosonic case.   It has been argued in \cite{titchmarsh2} that a pair of real-valued functions   $ ( A(z) \:\: B(z) )$  can be expanded   as a superposition of (\ref{eq:fermionicbasicfunctions1}) and   (\ref{eq:fermionicbasicfunctions2}) in the following form
\begin{eqnarray}
\left (
\begin{tabular}{c}
$A(z)$
\\
$B(z)$
\end{tabular}
\right )
&=&
 \sum _{k, l = 1,2}
\int 
\left (
\begin{tabular}{c}
$F_{k}^{(\pm)}(z; \Lambda)$
\\
$G_{k}^{(\pm)}(z; \Lambda)$
\end{tabular}
\right ) 
d\rho_{kl}^{(\pm)}(\Lambda)
\nonumber \\
& & \hskip1cm \times 
\int _{-\infty}^{+\infty} dz^{\:\prime} 
\left (
F_{l}^{(\pm)}(z^{\:\prime}; \Lambda )\: \; G_{l}^{(\pm)} (z^{\:\prime} ; \Lambda )
\right )
\left (
\begin{tabular}{c}
$A(z^{\:\prime })$
\\
$B(z^{\:\prime} )$
\end{tabular}
\right )\:.
\label{eq:eigenfunctionexpansionfermionic}
\end{eqnarray}
Here $d \rho^{(\pm)}_{kl} (\Lambda )$ describes the distribution of $\Lambda$   in connection  with the differential equation  (\ref{eq:standardform2}).   We use the letter $\Lambda $ instead of $E$ for the integration variable in (\ref{eq:eigenfunctionexpansionfermionic})  in order to avoid possible confusion that could perhaps arise later in (\ref{eq:EandLambdadiffer}).  According to  Ref. \cite{titchmarsh2}, the spectral functions are given by 
\begin{eqnarray}
d \rho_{11}^{(\pm)}(\Lambda_{}) 
&=&
\lim_{\varepsilon \to +0} \frac{1}{\pi } 
\Im \left (
\frac{1}{m^{(\pm)}_{2}(\Lambda + i \varepsilon)  - m^{(\pm)}_{1}(\Lambda + i \varepsilon) }
\right ) d\Lambda  \: ,
\label{eq:titchmarshkodaira4x}
\\
d \rho_{12}^{(\pm)}(\Lambda_{}) 
&=&
d \rho_{21}^{(\pm)}(\Lambda_{}) 
\nonumber \\
&=&
\lim_{\varepsilon \to +0} \frac{1}{2\pi } 
\Im \left ( 
\frac{ m^{(\pm)}_{2}(\Lambda + i \varepsilon) +  m^{(\pm)}_{1}(\Lambda + i \varepsilon) }
{m^{(\pm)}_{2}(\Lambda + i \varepsilon)  - m^{(\pm)}_{1}(\Lambda + i \varepsilon) }
\right ) d\Lambda \:,
\label{eq:titchmarshkodaira5x} 
\\
d \rho_{22}^{(\pm)}(\Lambda_{}) 
&=&
\lim_{\varepsilon \to +0} \frac{1}{\pi } 
\Im \left (
\frac{ m^{(\pm)}_{1}(\Lambda + i \varepsilon)  \:  m^{(\pm)}_{2}(\Lambda + 
i \varepsilon) }{m^{(\pm)}_{2}(\Lambda + i \varepsilon)  - m^{(\pm)}_{1}(\Lambda + i \varepsilon) }
\right ) d\Lambda \:,
\label{eq:titchmarshkodaira6x}
\end{eqnarray}
for $\Lambda > 0$, in case of continuous spectra.   (See Eq. (6.2) in Ref. \cite{titchmarsh2}. )
The characteristic functions $m_{1}^{(\pm)}(\Lambda + i \varepsilon )$ and $m_{2}^{(\pm)}(\Lambda 
+ i \varepsilon )$ are given by 
 \begin{eqnarray}
 m^{(\pm)}_{1}(\Lambda + i\varepsilon )
 &=& - \lim_{a \to  - \infty}\frac{F^{(\pm)}_{1}(a;  \Lambda + i \varepsilon )}
 {F^{(\pm)}_{2} ( a ;  \Lambda + i \varepsilon )}
 \:,
\nonumber  \\
  m^{(\pm)}_{2}(\Lambda + i\varepsilon )
&=&
 - \lim_{b \to  + \infty}\frac{F^{(\pm)}_{1}(b; \Lambda + i \varepsilon )}
  {F^{(\pm)}_{2} ( b ;  \Lambda + i \varepsilon  )}\:,
  \label{eq:characteristicfunctionfermion}
 \end{eqnarray}
 in a similar way as in the Bosonic case.  For $\Lambda <0$, the spectral functions are determined  by the symmetry property between  positive and negative energy solutions mentioned before. 

The  expansion  (\ref{eq:eigenfunctionexpansionfermionic}) may be equivalently expressed   by concise formulas 
\begin{eqnarray}
\sum_{k, l = 1,2}
\int F_{k}^{(\pm)}(z; \Lambda )\: d\rho_{kl}^{(\pm)} (\Lambda^{})\:F_{l}^{(\pm)}(z^{\:\prime}; \Lambda )&=& \delta (z - z^{\:\prime})\;,
\label{eq:conciseformulas1}
\\
\sum_{k, l = 1,2}
\int F_{k}^{(\pm)}(z; \Lambda )\: d\rho_{kl}^{(\pm)} (\Lambda^{})\:G_{l}^{(\pm)}(z^{\;\prime}; \Lambda )&=&0\;,
\label{eq:conciseformulas2}
\\
\sum_{k, l = 1,2}
\int G_{k}^{(\pm)}(z; \Lambda )\: d\rho_{kl}^{(\pm)} (\Lambda^{})\:F_{l}^{(\pm)}(z^{\;\prime}; \Lambda )&=&0\;,
\label{eq:conciseformulas3}
\\
\sum_{k, l  = 1,2}
\int G_{k}^{(\pm)}(z; \Lambda )\: d\rho_{kl }^{(\pm)} (\Lambda^{})\:G_{l}^{(\pm)}(z^{\:\prime}; \Lambda )
&=&\delta ( z- z^{\;\prime})\;.
\label{eq:conciseformulas4}
\end{eqnarray}
These formulas are further rephrased, through  the orthogonal rotation  like (\ref{eq:orthogonalrotation}), 
\begin{eqnarray}
\left (
\begin{tabular}{c}
$f_{k}^{(\pm)}(z; \Lambda  )$
\\
$g_{k}^{(\pm)}(z; \Lambda   )$
\end{tabular}
\right )
= 
\left (
\begin{tabular}{cc}
${\rm cos}\: \theta^{(\pm )}(z)$ & $ - {\rm sin}\:\theta^{ (\pm ) }(z) $
\\
$ {\rm sin}\:\theta^{ ( \pm )} (z) $ & ${\rm cos}\: \theta^{ ( \pm ) } (z) $
\end{tabular}
\right ) 
\left (
\begin{tabular}{c}
$F_{k}^{(\pm)}(z; \Lambda  )$
\\
$G_{k}^{(\pm)}(z; \Lambda  )$
\end{tabular}
\right )\:,
\hskip0.5cm
(k=1,2)\:,
\label{eq:orthogonalrotation2}
\nonumber \\
\end{eqnarray}
by the formulas
\begin{eqnarray}
\sum_{k, l = 1,2}
\int f_{k}^{(\pm)}(z; \Lambda )\: d\rho_{kl}^{(\pm)} (\Lambda^{})\:f_{l}^{(\pm)}(z^{\:\prime}; \Lambda )&=& \delta (z - z^{\:\prime})\;,
\label{eq:ffintegration}
\\
\sum_{k, l = 1,2}
\int f_{k}^{(\pm)}(z; \Lambda )\: d\rho_{kl}^{(\pm)} (\Lambda^{})\:g_{l}^{(\pm)}(z^{\;\prime}; \Lambda )&=&0\;,
\\
\sum_{k, l = 1,2}
\int g_{k}^{(\pm)}(z; \Lambda )\: d\rho_{kl}^{(\pm)} (\Lambda^{})\:f_{l}^{(\pm)}(z^{\;\prime}; \Lambda )&=&0\;,
\\
\sum_{k, l  = 1,2}
\int g_{k}^{(\pm)}(z; \Lambda )\: d\rho_{kl }^{(\pm)} (\Lambda^{})\:g_{l}^{(\pm)}(z^{\: \prime}; \Lambda )&=&\delta ( z- z^{\;\prime})\;.
\label{eq:ggintegration}
\end{eqnarray}

\subsection{Construction of the Propagator}

We are now in a position to construct the Fermion's two-point Green's function. 
Let us  write  down the solutions to the Dirac equation  (\ref{eq:eigenvalueproblemoriginal}) as
\begin{eqnarray}
\psi_{k}^{(\pm)} (z; \Lambda )  \equiv  \left (
\begin{tabular}{c}
  $f_{k}^{(\pm)} (z; \Lambda ) \eta^{(\pm)}$ 
  \\
  \\
    $g_{k}^{(\pm)} (z; \Lambda ) \eta^{(\pm)}$ 
\end{tabular}
\right ) \hskip1cm (k=1, 2)
\label{eq:1901}
\end{eqnarray}
and pay attention to  the following formula
\begin{eqnarray}
& & 
\sum _{k, l=1,2} \: \sum_{s=+, -} \int 
 \psi_{k}^{(s)}(z; \Lambda) d\rho ^{(s)}_{kl}(\Lambda^{})   \psi_{l}^{(s)}(z^{\:\prime}; \Lambda)^{\dag}
\nonumber \\
& & \hskip1cm =
\left (
\begin{tabular}{cc}
$\eta^{(+)} \eta^{(+) \dag }  + \eta^{(-)} \eta^{(-) \dag } $ & $0$
\\
$0$ & $\eta^{(+)} \eta^{(+) \dag }  + \eta^{(-)} \eta^{(-) \dag }  $
\end{tabular}
\right ) \delta(z - z^{\;\prime})
\nonumber \\
& & \hskip1cm =
\left (
\begin{tabular}{cc}
${\bf 1} $ & $0$
\\
$0$ & ${\bf 1}  $
\end{tabular}
\right ) \delta(z - z^{\;\prime})\:.
\label{eq:completenessrelation}
\end{eqnarray}
This formula can be derived by using (\ref{eq:ffintegration}) - (\ref{eq:ggintegration}) 
together with (\ref{eq:etaformula1}).
Noting that $\psi_{k}^{(\pm)}(z; \Lambda )$ satisfies the differential equation 
\begin{eqnarray}
\gamma^{0} \left \{
i \gamma \cdot \partial  - m(z)  \right \}  \psi_{k}^{(\pm )}(z; \Lambda ) 
e^{-iEt + i\vec{p}_{\perp} \cdot \vec{x}_{\perp}}
=
( E - \Lambda ) \psi _{k}^{(\pm)}(z; \Lambda )
e^{-iEt + i\vec{p}_{\perp} \cdot \vec{x}_{\perp}}\:, 
\label{eq:EandLambdadiffer}
\end{eqnarray}
we also find the formula
\begin{eqnarray}
& & \gamma^{0} \left \{ i \gamma \cdot \partial  - m(z) \right \}
\sum _{k, l=1,2} \: \sum _{s=+, -} \int 
\psi_{k}^{(s)}(z; \Lambda) \frac{d\rho ^{(s)}_{kl}(\Lambda^{})}{E-\Lambda^{}}   \psi_{l}^{(s)}(z^{\:\prime}; \Lambda)^{\dag}
e^{-iE ( t - t^{\prime} ) + i\vec{p}_{\perp} \cdot ( \vec{x}_{\perp} - \vec{x}_{\perp}^{\:\prime}  )}
\nonumber \\
& & \hskip3cm =
\left (
\begin{tabular}{cc}
${\bf 1} $ & ${\bf 0}$
\\
${\bf 0}$ & ${\bf 1}  $
\end{tabular}
\right ) \delta(z - z^{\;\prime})  e^{-iE ( t - t^{\prime} ) + i\vec{p}_{\perp} \cdot ( \vec{x}_{\perp} - \vec{x}_{\perp} )} \:.
\end{eqnarray}
By multiplying $\gamma^{0}$ from the left and another $\gamma^{0}$ from the right,  we get 
\begin{eqnarray}
& &  \left \{ i \gamma \cdot \partial  - m(z) \right \}
\int \frac{dE \: d^{2}\vec{p}_{\perp}}{(2\pi)^{3}}
 \: e^{-iE ( t - t^{\prime} ) + i\vec{p}_{\perp} \cdot ( \vec{x}_{\perp} - \vec{x}_{\perp}^{\:\prime}  )}
\nonumber \\
& & \hskip3cm
\times \sum _{k, l=1,2} \: \sum_{s=+, -} \int 
\psi_{k}^{(s)}(z; \Lambda) \frac{d\rho ^{(s)}_{kl}(\Lambda^{})}{E-\Lambda^{}}   \psi_{l}^{(s)}(z^{\:\prime}; \Lambda)^{\dag}\: \gamma^{0}
\nonumber \\
& & 
 \hskip7cm 
\nonumber \\
& & \hskip2cm =
\left (
\begin{tabular}{cc}
${\bf 1} $ & $0$
\\
$0$ & ${\bf 1}  $
\end{tabular}
\right ) \delta(z - z^{\;\prime})\delta(t-t^{\: \prime}) \delta^{2}(\vec{x}_{\perp} -\vec{x}_{\perp}^{\: \prime})\:.
\end{eqnarray}
We have thus  arrived finally at the integral representation for the Fermion's propagator 
\begin{eqnarray}
& & \hskip-0.2cm
{\cal F}(t-t^{\:\prime},  x - x^{\:\prime}, y - y^{\:\prime}, z, z^{\:\prime} ; m(*))
\nonumber \\
& &\hskip0cm =
i \int  \frac{dE \: d^{2} \vec{p}_{\perp}}{(2 \pi)^{3}} e^{-i E (t - t^{\:\prime})}
 e^{ i \vec{p}_{\perp} \cdot ( \vec{x}_{\perp}  - \vec{x}_{\perp}^{\:\prime})}   
\sum_{k,l=1,2} \: \sum_{s=+, -} 
 \int \psi_{k}^{(s)}(z; \Lambda ) \frac{d\rho^{(s)}_{kl} (\Lambda^{})}{E-\Lambda^{}}
\psi_{l}^{(s)}(z^{\:\prime}; \Lambda )^{\dag} \gamma^{0}\:.
\nonumber \\
\label{eq:fermionpropagatorfinalform}
\end{eqnarray}
In this expression we have not yet specified the $+i \varepsilon $ prescription in the energy denominator.  In order for the above Green's function to be the time-ordered vacuum expectation value of  the Fermion's bilinear operators we must go into the detail of the definition of the vacuum in the  presence of the bubble wall.  We will come to this point later in Sect.  \ref{sec:fieldquantization}, and for now suffice it to say that  the singularity at $E=\Lambda $ in (\ref{eq:fermionpropagatorfinalform}) should be avoided  by setting  
\begin{eqnarray}
\frac{d\rho^{(s)}_{kl} (\Lambda )}{E-\Lambda  }
\:\:\:\:  \to \:\:\:\:   
\frac{d\rho^{(s)}_{kl} (\Lambda )}{E-\Lambda + i \varepsilon } \:\:\:\:{\rm for}\:\:\:\: \Lambda > 0\:,
\label{eq:prescription1}
\end{eqnarray}
and
\begin{eqnarray}
\frac{d\rho^{(s)}_{kl} (\Lambda )}{E-\Lambda  }
\:\:\:\:  \to \:\:\:\:    
\frac{d\rho^{(s)}_{kl} (\Lambda )}{E-\Lambda - i \varepsilon } \:\:\:\:{\rm for}\:\:\:\: \Lambda < 0\:.
\label{eq:prescription2}
\end{eqnarray}

 \subsection{The constant mass case (no bubble wall)   }
 
 Let us put our formula (\ref{eq:fermionpropagatorfinalform}) into an explicit one for the case that the mass is just a constant, i.e., $m(z)=m$. In this case $Q(z)$ in (\ref{eq:defofq(z)}) is just a constant $\sqrt{m^{2} + \vert \vec{p}_\perp \vert^{2} }$ and it is a simple matter to get the solutions to   (\ref{eq:standardform2}).   For $\Lambda ^{2} > ( m^{2} + \vert \vec{p}_{\perp}\vert ^{2})$,  the solutions become 
  \begin{eqnarray}
 & &  F_{1}^{(\pm)}(z; \Lambda)={\rm cos}\:(qz),
 \;\;\;\:\:\:\:
 G_{1}^{(\pm)}(z;  \Lambda)=\frac{ \Lambda^{} \pm  \sqrt{m^{2}+  \vert \vec{p}_{\perp} \vert^{2}}}{q}
 {\rm sin}\:(qz)\:,
 \label{eq:solution354}
 \\
 & &
 F_{2}^{(\pm)}(z; \Lambda)= - \frac{ \Lambda^{} \mp  \sqrt{m^{2}+  \vert \vec{p}_{\perp} \vert^{2}}}{q}
 {\rm sin}\:(qz)\:, 
 \;\;\;\;
 G_{2}^{(\pm)}(z; \Lambda) = {\rm cos}\:(qz)\:,
 \label{eq:solution355}
 \end{eqnarray}
 where we defined $q$ as follows
 \begin{eqnarray}
 q=\sqrt{\Lambda^{2} - ( m^{2} + \vert \vec{p}_{\perp} \vert ^{2} ) }\:.
 \label{eq:definitionofq}
 \end{eqnarray}
 For $\Lambda ^{2} < (m^{2} + \vert \vec{p}_{\perp}\vert ^{2}) $, on the other hand, the solutions are 
 \begin{eqnarray}
 & &  F_{1}^{(\pm)}(z; \Lambda)={\rm cosh}\:( {\tilde q}z),
 \;\;\;\:\:\:\:
 G_{1}^{(\pm)}(z;\Lambda)=\frac{ \Lambda^{} \pm  \sqrt{m^{2}+  \vert \vec{p}_{\perp} \vert^{2}}}{
 {\tilde q}}
 {\rm sinh}\:( {\tilde q} z)\:,
 \label{eq:solution357}
 \\
 & &
 F_{2}^{(\pm)}(z;\Lambda)= - \frac{ \Lambda^{} \mp  \sqrt{m^{2}+  \vert \vec{p}_{\perp} \vert^{2}}}{
 {\tilde q}}
 {\rm sinh}\:( {\tilde q} z)\:, 
 \;\;\;\;
 G_{2}^{(\pm)}(z; \Lambda) = {\rm cosh}\:( {\tilde q} z)\:.
 \label{eq:solution358}
 \end{eqnarray}
 Here $\tilde{q}$ is defined  by 
 \begin{eqnarray}
{\tilde  q} =\sqrt{ ( m^{2} + \vert \vec{p}_{\perp} \vert ^{2} )
-   \Lambda^{2}
 }\:.
 \end{eqnarray}

 The spectral functions  (\ref{eq:titchmarshkodaira4x}) -  (\ref{eq:titchmarshkodaira6x}), are also explicitly computed by using the formula (\ref{eq:characteristicfunctionfermion}) as follows, 
 \begin{eqnarray}
d\rho_{11}^{(\pm)} (\Lambda ^{})&=&
 \left \{
 \begin{tabular}{l}
 $
 \displaystyle{
 \frac{1}{2\pi}\frac{\Lambda^{}  \mp  \sqrt{m^{2} + \vert \vec{p}_{\perp} \vert^{2}}}
 {q}\: d\Lambda 
 }
 $
 \hskip0.7cm 
 {\rm for} \:\:\:  $\Lambda ^{} >  \sqrt{ m^{2} + \vert \vec{p}_{\perp}\vert ^{2} }  $\:,
 \\
 $0$  \hskip4.6cm 
 {\rm for} \:\:\:  $\Lambda ^{2} <  ( m^{2} + \vert \vec{p}_{\perp}\vert ^{2}) $\:,
 \\
$
 \displaystyle{
 \frac{1}{2\pi}\frac{ - \Lambda^{}  \pm  \sqrt{m^{2} + \vert \vec{p}_{\perp} \vert^{2}}}
 {q}\: d\Lambda 
 }
 $
 \hskip0.5cm 
 {\rm for} \:\:\:  $\Lambda  <  - \sqrt{ m^{2} + \vert \vec{p}_{\perp}\vert ^{2} }  $\:,
\end{tabular}
 \right.
 \\
 \nonumber \\
 d\rho_{12}^{(\pm)} (\Lambda ^{})&=& d\rho_{21}^{(\pm)} (\Lambda ^{})=0\:,
 \\
 \nonumber \\
d\rho_{22}^{(\pm)} (\Lambda ^{})&=& 
 \left \{
 \begin{tabular}{l}
 $
 \displaystyle{
 \frac{1}{2\pi}\frac{\Lambda^{}  \pm  \sqrt{m^{2} + \vert \vec{p}_{\perp} \vert^{2}}}
 {q} \: d\Lambda
 }
 $
 \hskip0.7cm 
 {\rm for} \:\:\:  $ \Lambda ^{} >  \sqrt{ m^{2} + \vert \vec{p}_{\perp}\vert ^{2} }$\:,
 \\
 $0$  \hskip4.6cm 
 {\rm for} \:\:\:  $ \Lambda^{2} < (m^{2} + \vert \vec{p}_{\perp}\vert ^{2})$\:,
 \\
 $
 \displaystyle{
 \frac{1}{2\pi}\frac{ - \Lambda^{}  \mp  \sqrt{m^{2} + \vert \vec{p}_{\perp} \vert^{2}}}
 {q}\: d\Lambda 
 }
 $
 \hskip0.5cm 
 {\rm for} \:\:\:  $\Lambda  <  - \sqrt{ m^{2} + \vert \vec{p}_{\perp}\vert ^{2} }  $\:.
\end{tabular}
 \right.
 \end{eqnarray}
 Here  we notice a simple relation   $d\rho_{11}^{(\pm)}(-\Lambda ) / d(- \Lambda ) =  d\rho_{22}^{(\pm)}(\Lambda  ) / d\Lambda $   which is  due to the positive- and negative energy correspondence.  With these spectral functions we can  confirm the completeness relations    (\ref{eq:conciseformulas1}) - (\ref{eq:conciseformulas4}) explicitly.
 
 In order to  put the formula (\ref{eq:fermionpropagatorfinalform}) into a more familiar form, we have to use the solutions (\ref{eq:solution354})  and  (\ref{eq:solution355}) together with the spectral functions to get the integrations 
\begin{eqnarray}
& & 
\sum_{k, l = 1,2}
\int F_{k}^{(\pm)}(z; \Lambda )\: \frac{d\rho_{kl}^{(\pm)} (\Lambda^{})}{E-\Lambda } 
\:F_{l}^{(\pm)}(z^{\:\prime}; \Lambda )
\nonumber \\
& &\hskip2cm 
=
\frac{1}{2\pi} \int _{-\infty}^{+\infty}\: dq\: e^{iq (z - z^{\:\prime})}
\frac{E \mp \sqrt{m^{2} + \vert \vec{p}_{\perp} \vert^{2}}}
{E^{2} - (m^{2} + \vert \vec{p}_{\perp} \vert^{2} + q^{2}) + i \varepsilon }
\;,
\end{eqnarray}
\begin{eqnarray}
& & 
\sum_{k, l = 1,2}
\int F_{k}^{(\pm)}(z; \Lambda )\: \frac{d\rho_{kl}^{(\pm)} (\Lambda^{})}{E- \Lambda }
\:G_{l}^{(\pm)}(z^{\;\prime}; \Lambda )
\nonumber \\
& & \hskip2cm =\frac{1}{2\pi} \int _{-\infty}^{+\infty}\: dq\: e^{iq (z - z^{\:\prime})}
\frac{iq}{E^{2} - (m^{2} + \vert \vec{p}_{\perp} \vert^{2} + q^{2}) + i \varepsilon}
\;,
\end{eqnarray}
\begin{eqnarray}
& & 
\sum_{k, l = 1,2}
\int G_{k}^{(\pm)}(z; \Lambda )\: \frac{d\rho_{kl}^{(\pm)} (\Lambda^{})}{E-\Lambda}
\:F_{l}^{(\pm)}(z^{\;\prime}; \Lambda )
\nonumber \\
& &  \hskip2cm 
=\frac{1}{2\pi} \int _{-\infty}^{+\infty}\: dq\: e^{iq (z - z^{\:\prime})}
\frac{- iq}{E^{2} - (m^{2} + \vert \vec{p}_{\perp} \vert^{2} + q^{2}) + i \varepsilon}
  \;,
\end{eqnarray}
\begin{eqnarray}
& & 
\sum_{k, l  = 1,2}
\int G_{k}^{(\pm)}(z; \Lambda )\: \frac{d\rho_{kl }^{(\pm)} (\Lambda^{})}{E-\Lambda} 
\:G_{l}^{(\pm)}(z^{\: \prime}; \Lambda )
\nonumber \\
& & \hskip2cm 
=
\frac{1}{2\pi} \int _{-\infty}^{+\infty}\: dq\: e^{iq (z - z^{\:\prime})}
\frac{E \pm \sqrt{m^{2} + \vert \vec{p}_{\perp} \vert^{2}}}{E^{2} - (m^{2} + \vert \vec{p}_{\perp} \vert^{2} + q^{2}) + i \varepsilon}
  \;.
\end{eqnarray}
According to the original definition (\ref{eq:definitionofq}),   $q$ is positive,   but the integration region of $q$ in the above formulas  is extended to $( -\infty , +\infty )$ by using symmetry arguments.  These formulas may be transformed via orthogonal transformation (\ref{eq:orthogonalrotation2}) into the  formulas involving  $f_{k}^{(\pm)} (z; \Lambda )$ and $g_{k}^{(\pm)} (z; \Lambda )$, 
\begin{eqnarray}
& & 
\sum_{k, l = 1,2}
\int f_{k}^{(\pm)}(z; \Lambda )\: \frac{d\rho_{kl}^{(\pm)} (\Lambda^{})}{E-\Lambda } 
\:f_{l}^{(\pm)}(z^{\:\prime}; \Lambda )
\nonumber \\
& &\hskip2cm 
=
\frac{1}{2\pi} \int _{-\infty}^{+\infty}\: dq\: e^{iq (z - z^{\:\prime})}
\frac{E \mp
 \vert \vec{p}_{\perp} \vert }{E^{2} - (m^{2} + \vert \vec{p}_{\perp} \vert^{2} + q^{2}) + i \varepsilon}
\;,
\end{eqnarray}
\begin{eqnarray}
& & 
\sum_{k, l = 1,2}
\int f_{k}^{(\pm)}(z; \Lambda )\: \frac{d\rho_{kl}^{(\pm)} (\Lambda^{})}{E- \Lambda }
\:g_{l}^{(\pm)}(z^{\;\prime}; \Lambda )
\nonumber \\
& & \hskip2cm =\frac{1}{2\pi} \int _{-\infty}^{+\infty}\: dq\: e^{iq (z - z^{\:\prime})}
\frac{m + iq}{E^{2} - (m^{2} + \vert \vec{p}_{\perp} \vert^{2} + q^{2}) + i \varepsilon}
\;,
\end{eqnarray}
\begin{eqnarray}
& & 
\sum_{k, l = 1,2}
\int g_{k}^{(\pm)}(z; \Lambda )\: \frac{d\rho_{kl}^{(\pm)} (\Lambda^{})}{E-\Lambda}
\:f_{l}^{(\pm)}(z^{\;\prime}; \Lambda )
\nonumber \\
& &  \hskip2cm 
=\frac{1}{2\pi} \int _{-\infty}^{+\infty}\: dq\: e^{iq (z - z^{\:\prime})}
\frac{ m - i q}{E^{2} - (m^{2} + \vert \vec{p}_{\perp} \vert^{2} + q^{2}) + i \varepsilon}
  \;,
\end{eqnarray}
\begin{eqnarray}
& & 
\sum_{k, l  = 1,2}
\int g_{k}^{(\pm)}(z; \Lambda )\: \frac{d\rho_{kl }^{(\pm)} (\Lambda^{})}{E-\Lambda} 
\:g_{l}^{(\pm)}(z^{\: \prime}; \Lambda )
\nonumber \\
& & \hskip2cm 
=
\frac{1}{2\pi} \int _{-\infty}^{+\infty}\: dq\: e^{iq (z - z^{\:\prime})}
\frac{ E \pm \vert \vec{p}_{\perp} \vert }{E^{2} - (m^{2} + \vert \vec{p}_{\perp} \vert^{2} + q^{2}) 
+ i \varepsilon}
  \;.
\end{eqnarray}

Combining all these formulas,    we end up with 
\begin{eqnarray}
& &
 \sum_{k,l=1,2} \: \sum_{s=+, -} 
 \int \psi_{k}^{(s)}(z; \lambda ) \frac{d\rho^{(s)}_{kl} (\Lambda^{})}{E-\Lambda^{}}
\psi_{l}^{(s)}(z^{\:\prime}; \lambda )^{\dag} \gamma^{0}
   \nonumber \\
   & &\hskip2cm 
=
\frac{1}{2\pi} \int _{-\infty}^{+\infty}  dq\:e^{iq(z-z^{\:\prime})}\: 
\frac{E\gamma^{0} - \gamma^{1}p_{x} -\gamma^{2}p_{y} -q\gamma^{3} +m}
{E^{2} - \vert \vec{p}_{\perp}\vert^{2} - q^{2} -m^{2} + i \varepsilon}\:.
\label{eq:almostfinished}
\end{eqnarray}
Here use has been made  of the identities  (\ref{eq:etaformula1}) and (\ref{eq:etaformula2}). 
Note, in particular, that the momenta $p_{x}$ and $p_{y}$ 
in the numerator  of (\ref{eq:almostfinished}) show up through  the use of 
the formula (\ref{eq:etaformula2}). 
Putting the above relation  into (\ref{eq:fermionpropagatorfinalform}), we finally arrive at 
the familiar propagator  (\ref{eq:constantmassfermion}) with $p^{\mu}=(E, p_{x}, p_{y}, q)$.  
We have thus successfully reproduced the well-known formula with the aid of less-known method using 
 the spectral function $\rho_{kl}(\Lambda )$. In the presence of the bubble wall, 
it seems  absolutely necessary to make a full use of the spectral function to construct Green's functions.

\section{The vector boson's two-point Green's function}

\subsection{Mixing between gauge boson and unphysical scalar fields}
Now let us turn to the Green's functions of the gauge bosons, in which we encounter with  some peculiar phenomena in contrast to the scalar case. The polarization vectors of the vector bosons are usually defined referring  to the four-momentum, whose $z$-component in the present case, however, is changing  during the passage  of the bubble wall. The separation into transverse and longitudinal polarization states becomes therefore more  involved than usual. Moreover the unphysical scalar fields are also mixed up with the vector bosons through the variation of the Higgs vacuum expectation value. Farrar and  McIntosh \cite{farrar}  once discussed the change of the polarization states that   vector bosons meet with in the course of passing through the bubble wall. Azatov et al also discussed in their work \cite{azatov4} the alteration of the polarization vector by choosing the unitary gauge. We will go along similar lines as theirs by making use of the 't Hooft-Feynman gauge  which is more easy to deal with to construct our propagators.

Let us consider specifically the standard $SU(2)_{L} \times U(1)_{Y}$ electroweak theory in order to see the mixing between the unphysical scalar and gauge boson fields. The  Higgs Lagrangian is  given as usual by 
 \begin{eqnarray}
 {\cal L}_{\rm scalar}&=&( \nabla_{\mu} \Phi )^{\dag} (\nabla^{\mu} \Phi ) - V(\Phi)\:,
 \hskip1cm 
 V(\Phi )= \lambda (\Phi ^{\dag} \Phi )^{2} + \mu_{0}^{2} \: \Phi ^{\dag} \Phi \:,
 \label{eq:scalarlagrangian}
 \end{eqnarray}
 where $\nabla_{\mu}$ is the gauge covariant derivative and the Higgs doublet field $\Phi$ is parametrized as 
 \begin{eqnarray}
 \Phi = \left (
 \begin{tabular}{c}
 $w^{\dag}$
 \\
\\
 $\displaystyle{\frac{1}{\sqrt{2}}(v + H + i \chi )}$
 \end{tabular}
 \right )\:.
 \label{eq:higgsdoublet}
 \end{eqnarray}
 The physical Higgs field is denoted by $H$ and $w$, $w^{\dag}$ and $z$ are the unphysical scalar fields. In contrast with the usual case, the vacuum expectation value $v$ is not  a constant but is supposed to be position dependent,  i.e., $\partial_{\mu}v \neq 0$.  Putting (\ref{eq:higgsdoublet}) into (\ref{eq:scalarlagrangian}) and keeping terms up to the quadratic ones, we get
 \begin{eqnarray}
 {\cal L}_{\rm scalar}
 &=&
 \partial_{\mu}w^{\dag}\partial^{\mu}w 
 + \frac{1}{2}\partial_{\mu} (v+H-i \chi ) \partial^{\mu} (v + H +i \chi ) 
 \nonumber   \\
& &  
- M_{Z} Z^{\mu} \left ( \partial _{\mu} \chi  -  \chi \frac{ \partial_{\mu} v }{v}  \right )
\nonumber 
\\
& & +i M_{W}  W^{+ \mu} \left ( \partial_{\mu} w -   w \frac{ \partial_{\mu} v }{v}    \right )
- i M_{W}   W^{- \mu} \left ( \partial_{\mu} w^{\dag}  - w^{\dag} \frac{ \partial_{\mu} v }{v}   \right )
\nonumber 
\\
& & + M_{W}^{2} W_{\mu}^{+}W^{- \mu}   +   \frac{1}{2}M_{Z}^{2} Z_{\mu}Z^{\mu}  
\nonumber 
\\
& & + (\lambda v^{2}   + \mu_{0} ^{2} ) \left (w^{\dag}w + \frac{1}{2} \chi ^{2} \right )
+ \left ( \frac{3}{2} \lambda v^{2} + \frac{1}{2} \mu_{0}^{2} \right ) H^{2}
+ \left ( \lambda v^{3} + v \mu_{0}^{2} \right ) H
\nonumber \\
& &     + \cdots \cdots \:,
\label{eq:quadraticscalarlagrangian}
 \end{eqnarray}
 where  ellipses denote cubic and quartic terms with respect to fields. Note that the  $W$ and $Z$   boson masses 
  \begin{eqnarray}
 M_{W}^{2}=\frac{1}{4}g^{2} v^{2}, 
 \hskip1cm
 M_{Z} ^{2} = \frac{1}{4}( g^{2} + g^{ \prime \: 2} ) v^{2} 
 \end{eqnarray}
are also position dependent.    Looking at (\ref{eq:quadraticscalarlagrangian}), we immediately notice  that there are extra terms 
\begin{eqnarray}
\left (
 M_{Z} Z^{\mu}  \chi 
 - i M_{W} W^{+ \mu}  w 
+ i M_{W} W^{- \mu}  w^{\dag}
\right )   \frac{\partial _{\mu}v}{v}  
\label{eq:mixingterms}
\end{eqnarray}
 for $\partial_{\mu} v \neq 0$   in comparison with the conventional case.  This describes the mixing between gauge bosons and associated unphysical scalar fields.  

 In the following we would like to employ the 't Hooft-Feynman ($\xi=1$ in $R_{\xi}$) gauge and add the following gauge fixing (GF) Lagrangian 
 \begin{eqnarray}
 {\cal L}_{\rm GF}
 &=& 
 -\big (  \partial \cdot W_{}^{+} + i   M_{W} w ^{\dag}   \big )
 \big (  \partial \cdot W_{}^{-} - i    M_{W}  w    \big )
 -\frac{1}{2}\big (
 \partial \cdot  Z_{}   +     M_{Z} \:  \chi
 \big )^{2}
 -\frac{1}{2} \left (  \partial \cdot  A_{} \right )^{2}\:.
 \nonumber \\
 \label{eq:gaugefixing}
 \end{eqnarray}
(The gauge fixing term for the electromagnetic field $A_{\mu}$ is included in (\ref{eq:gaugefixing})  for completeness,  but will  be irrelevant to the subsequent analyses.)   The $R_{\xi}$ gauge is usually arranged to eliminate the mixing terms between gauge boson and associated unphysical scalar fields.  In the present case, however, the mixing does not disappear if  $\partial _{\mu} v \neq 0$. In fact,  in the  partial integration in (\ref{eq:gaugefixing}) we  use  the relations
\begin{eqnarray}
\partial_{\mu} M_{W}=M_{W} \frac{\partial _{\mu} v}{v}, 
\hskip1cm
\partial_{\mu} M_{Z}=M_{Z} \frac{\partial _{\mu} v}{v}, 
\end{eqnarray}
 and we notice that the mixing terms in (\ref{eq:mixingterms}) are not cancelled but are doubled. The free equations of motion of the $W$- and $Z$-bosons and associated unphysical scalars turn out to be  
 \begin{eqnarray}
& &   \left (  \square + M_{Z}^{2} \right )   Z_{\mu}   + 2M_{Z}  \frac{ \partial _{\mu} v }{v}  \chi =0\:,
    \\
& &   \left (  \square + M_{W}^{2} \right )   W^{-}_{\mu}   -  2i M_{W} \frac{ \partial _{\mu} v }{v}  w =0   \:,
 \hskip0.3cm
  \left (  \square + M_{W}^{2} \right )  W^{+}_{\mu}   +  2i M_{W} \frac{ \partial _{\mu} v }{v}  w^{\dag } =0   \:,
\\
& &  \left (     \square + M_{1}^{2} \right )  \chi  -   2M_{Z} \frac{\partial_{\mu} v}{v}  Z^{\mu}
 =0 \:,
 \\
& &   \left (     \square + M_{2}^{2} \right )  w -   2i M_{W} \frac{\partial_{\mu} v}{v}  W^{- \mu}
 =0 \:,
\hskip0.3cm
  \left (     \square + M_{2}^{2} \right )  w^{\dag } +   2i M_{W} \frac{\partial_{\mu} v}{v}  W^{+ \mu}
 =0 \:,
  \end{eqnarray}
  and show mixing apparently,    where use has been made of the notations
 \begin{eqnarray}
 M_{1}^{2}   =  M_{Z}^{2} + \left ( \lambda v^{2} + \mu_{0}^{2} \right )\:,
 \hskip1cm
  M_{2}^{2}   =  M_{W}^{2} + \left ( \lambda v^{2} + \mu_{0}^{2} \right )\:.
  \label{eq:m1m2}
 \end{eqnarray}
 Note that the term $\lambda v^{2} + \mu_{0}^{2}$ in (\ref{eq:m1m2}) goes to zero in the symmetry-broken region, but we have to keep it in  general.
 
 The mixing is, however, rather limited, if $v$ depends on the $z$-coordinate only, i.e., $v=v(z)$.  In fact 
 the components of vector fields with $\mu=0,1,2$ are free from mixing
 \begin{eqnarray}
   \left (  \square + M_{Z}^{2} \right )   Z^{\: i}=0\:,
    \hskip1cm 
  \left (  \square + M_{W}^{2} \right )   W^{\pm \: i} =0   \:,
  \hskip1cm (i=0,1,2)\:,
 \end{eqnarray}
 and the two-point Green's functions $\langle ( {\rm T} (Z^{i}Z^{j}) \rangle $ and 
 $\langle {\rm T} (W^{\pm \: i }W^{\mp \: j}) \rangle $ $(i, j =0,1,2)$ are easily worked out in the same way as in the scalar field case described in Sect. \ref{sec:green}.
 The mixing occurs only in the third components,  $Z^{3}$ and $W^{\pm \: 3}$, with their respective unphysical scalar fields.  The  equations of motion for the pair   $( Z^{3} \:\:\chi )$ become
 \begin{eqnarray}
 \left (  \square + M_{Z}^{2} \right )   Z^{3}   - 2M_{Z}  \frac{  v^{\prime}(z) }{v(z)}  z =0\:,
 \hskip0.3cm
 \left (     \square + M_{1}^{2} \right )  \chi -   2M_{Z} \frac{ v^{\prime}(z) }{v(z) }  Z^{3}
 =0 \:,
 \label{eq:z3chifreeequation}
 \end{eqnarray}
 and similar equations can be written down for the pairs,   $( W^{-\:3} \:\:\: w)$ and  $( W^{+\:3} \:\:\: w^{\dag })$ as follows
  \begin{eqnarray}
 \left (  \square + M_{W}^{2} \right )   W^{- 3}   +  2i M_{W}  \frac{  v^{\prime}(z) }{v(z)}  w =0\:,
 \hskip0.3cm
 \left (     \square + M_{2}^{2} \right )  w -   2i M_{W} \frac{ v^{\prime}(z) }{v(z) }  W^{-3} =0 \:,
 \end{eqnarray}
  \begin{eqnarray}
 \left (  \square + M_{W}^{2} \right )   W^{+ 3}   -  2i M_{W}  \frac{  v^{\prime}(z) }{v(z)}  w^{\dag} =0\:,
 \hskip0.3cm
 \left (     \square + M_{2}^{2} \right )  w^{\dag }  +   2i M_{W} \frac{ v^{\prime}(z) }{v(z) }  W^{+3} =0 \:.
 \end{eqnarray}

 \subsection{The general expansion method for two-component fields} 
 
 Since the third component of the gauge boson is mixed with the associated unphysical scalar boson, 
 the two-point function that we are seeking for becomes necessarily a $2 \times 2$ matrix and let us explain what the matrix looks like  for the pair $( Z^{3} \:\:\: \chi )$.  (The case of the $W$ boson and $w$ can be handled  exactly in the same way.)

 Looking at Eq. (\ref{eq:z3chifreeequation}), we notice that the $2 \times 2$ Green's function matrix has to satisfy the following equation, 
 \begin{eqnarray}
 & &
 \left (
 \begin{tabular}{cc}
 $\square +M_{Z}^{\:2}$ & $\displaystyle{- 2M_{Z} \frac{v^{\prime}(z)}{v(z)}}$
 \\
 \\
 $\displaystyle{- 2M_{Z} \frac{v^{\prime}(z)}{v(z)}}$ & $\square +M_{1}^{\:2}$
 \end{tabular}
 \right )
 \left (
 \begin{tabular}{cc}
 ${\cal G}_{11}$ & ${\cal G}_{12}$
 \\
 \\
 ${\cal G}_{21}$ & ${\cal G}_{22}$
 \end{tabular}
 \right )
 \nonumber  \\
& & \hskip2cm
 =-i \left (
 \begin{tabular}{cc}
 $1$ & $0$
 \\
 $0$ & $1$
 \end{tabular}
 \right )\delta (t-t^{\:\prime}) \delta (x - x^{\:\prime}) \delta (y - y^{\:\prime} )\delta  (z-z^{\:\prime})\:.
 \end{eqnarray}
 The translation invariance in time and the transverse $\vec{x}_{\perp}$-directions is respected and each component of the matrix is a function of the following variables, 
 \begin{eqnarray}
 {\cal G}_{\alpha \beta} =  {\cal G}_{\alpha \beta}( t-t^{\:\prime} , x - x^{\:\prime}, y-y^{\:\prime}, z, z^{\:\prime}; v(*)), 
 \hskip1cm
 (\alpha, \beta = 1,2))\:,
 \end{eqnarray}
 and is  expressed as a superposition of the plane wave
 \begin{eqnarray}
 e^{-iE(t-t^{\prime})}e^{i\vec{p}_{\perp} \cdot ( \vec{x}_{\perp} - \vec{x}_{\perp}^{\:\prime})}\:.
 \end{eqnarray}
The d'Alembertian  is therefore going to be replaced as
 \begin{eqnarray}
 \square \to -E^{2} +\vec{p}_{\perp}^{\: 2} - \frac{\partial^{2}}{\partial z^{2}}
 \end{eqnarray}
 and we are naturally led to  consider the differential equation of the following type
 \begin{eqnarray}
 \bigg \{
 -\frac{d^{2}}{dz^{2}} + V(z)  \bigg \} \phi (z; \lambda )= \lambda \phi (z; \lambda )\:.
 \label{eq:vectorscalareigenvalueproblem}
 \end{eqnarray}
 This is again the Schr{\" o}dinger-type equation with the ``potential''  $V(z)$, which is a  $2 \times 2 $ matrix defined by 
 \begin{eqnarray}
 V(z)=
 \left (
 \begin{tabular}{cc}
 $M_{Z}^{\:2}$ & $\displaystyle{- 2M_{Z} \frac{v^{\prime}(z)}{v(z)}}$
 \\
 \\
 $\displaystyle{- 2M_{Z} \frac{v^{\prime}(z)}{v(z)}}$ & $M_{1}^{\:2} $
 \end{tabular}
 \right )\:,
 \label{eq:potentialmatrixv(z)}
 \end{eqnarray}
 The ``energy''  is denoted by  $\lambda$  in ( \ref{eq:vectorscalareigenvalueproblem}), and $\phi (z; \lambda ) $ is the corresponding wave function with two components
 \begin{eqnarray}
 \phi (z; \lambda ) = \left (
 \begin{tabular}{c}
 $\phi_{1}(z; \lambda )$
 \\
\\
 $\phi_{2} (z;\lambda )$
 \end{tabular}
 \right )\:.
 \end{eqnarray}
Note that the matrix   (\ref{eq:potentialmatrixv(z)}) is real-symmetric, and the operator on the left hand side of  (\ref{eq:vectorscalareigenvalueproblem}) is   self-adjoint  in the space of functions with favorable behaviors at $ z \to \pm \infty$.
 The differential  equation (\ref{eq:vectorscalareigenvalueproblem}) can be rewritten equivalently in the following form, 
 \begin{eqnarray}
 & &
 \left (
 \begin{tabular}{cc}
 $\square +M_{Z}^{\:2}$ & $\displaystyle{- 2M_{Z} \frac{v^{\prime}(z)}{v(z)}}$
 \\
 \\
 $\displaystyle{- 2M_{Z} \frac{v^{\prime}(z)}{v(z)}}$ & $\square +M_{1}^{\:2}$
 \end{tabular}
 \right )\phi (z;\lambda )e^{-iEt} e^{i \vec{p}_{\perp} \cdot \vec{x}_{\perp} }
  \nonumber \\
 & &  \hskip4cm 
= (-E^{2} + \vec{p}_{\perp}^{\: 2} + \lambda ) 
 \phi (z;\lambda )e^{-iEt} e^{i \vec{p}_{\perp} \cdot \vec{x}_{\perp} }\:.
   \end{eqnarray}

 At this point let us recall that Kodaira,   in his follow-up paper \cite{kodaira2},   extended  the work  done in  \cite{weyl},  \cite{stone},  \cite{titchmarsh} and  \cite{kodaira1} on the general expansion method    associated with the second-order differential operator  to the case of any even-order  self-adjoint differential operator.  As remarked almost at  the end of Ref.   \cite{kodaira2},   the  expansion method developed there   can be   carried  to the case of coupled differential equation of lower-order for multi-component fields and our present problem belongs to such a category of the two-component case. 
 
 We can proceed similarly as in Sect.  \ref{sec:green} and  Sect. \ref{sec:fermiongreenfunction}
 by preparing  fundamental solutions to (\ref{eq:vectorscalareigenvalueproblem}), which we denote 
 $\phi^{(k)} (z; \lambda )$ $\:\:(k=1, \cdots , 4)$.  They  are defined by the initial conditions at $z=0$. 
 All the solutions to the Schr{\" o}dinger equation  (\ref{eq:vectorscalareigenvalueproblem}) are expressed as  a linear combination of $\phi^{(k)}(z; \lambda )$.  It has been shown that there exists the spectral function,  $\rho_{kl}(\lambda ) $with such a property that any pair of square-integrable functions 
 $u_{\alpha }(z) (\alpha =1,2)$, can be expressed as a superposition of the  fundamental solutions   
 \begin{eqnarray}
 u_{\alpha} (z)=\int \sum_{k,l} \phi^{(k)}_{\alpha} (z; \lambda ) d\rho_{kl}(\lambda )
 \int _{-\infty}^{+\infty} dz^{\:\prime}  \sum_{\beta =1,2} \phi_{\beta}^{(l)}(z^{\:\prime} \; \lambda )u_{\beta}(z^{\: \prime})\:.
 \end{eqnarray}
 This indicates that we may rewrite this formula in a simple form as
 \begin{eqnarray}
 \int \sum _{k,l } \phi_{\alpha} ^{(k)}(z; \lambda ) d\rho_{k l}(\lambda)  \phi_{\beta} ^{(l)}(z^{\prime}; \lambda )
 =
 \delta _{\alpha \beta} \: \delta (z - z^{\prime})\:.
 \label{eq:twocomponentcompleteness}
 \end{eqnarray}
 With the help of  (\ref{eq:twocomponentcompleteness}), it is almost straightforward to write down the Green's function in the following integral form
\begin{eqnarray}
& & \hskip-1cm
  {\cal G}_{\alpha \beta}( t-t^{\:\prime} , x - x^{\:\prime}, y-y^{\:\prime}, z, z^{\:\prime}; v(*))
  \\\nonumber \\
 &=&
i  \int \frac{dE \: d^{2}\vec{p}_{\perp}}{(2\pi)^{3}}  
 e^{-i E (t-t^{\:\prime})} e^{i \vec{p}_{\perp} \cdot 
 (\vec{x}_{\perp} - \vec{x}_{\perp}^{\:\prime})} 
\int \sum_{k,l}
 \frac{\phi^{(k)}_{\alpha}(z; \lambda) d\rho_{kl}(\lambda )  \phi^{(l)}_{\beta} (z^{\:\prime} ; \lambda ) }{E^{2} - \vec{p}_{\perp}^{\:2} - \lambda + i \varepsilon }\:.
 \label{eq:twocomponentgreensfunction}
 \end{eqnarray}
 We have added $+i\varepsilon$ in the denominator in (\ref{eq:twocomponentgreensfunction}) just by hand so that this should be the vacuum expectation value of the time-ordered product of fields in the sense to be explained next.

 \section{Field quantization in the presence of the bubble wall}
 \label{sec:fieldquantization}
 
In  constructing the two-point functions, we  made use  of  various propagation modes of    scalar and vector fields labelled by $\lambda$,  and of spinor fields labelled by $\Lambda$.   We are now well-prepared to discuss how to quantize these  fields by taking the existence of   the bubble wall into account.  For the sake of simplicity we consider the case in which the spectral functions admit   only continuous spectra. We will discuss below the quantization of scalar and spinor fields, while the vector fields will be dealt with  in  our separate publication since the physical state conditions are complicated and we have to prepare other tools to explain them.
 
  \subsection{Real scalar field }

 Let us begin with the scalar field satisfying the free equations of motion
 \begin{eqnarray}
\left \{   \square  + M(z)^{2} \right \}   \varphi (t, \vec{x}_{\perp}, z)  = 0\:,
\label{eq:equationsofmotion}
 \end{eqnarray}
  in the presence of the bubble wall.  Our task now is to realize the canonical commutation relations 
\begin{eqnarray}
& & 
\left [ \dot{ \varphi } (t, \vec{x}_{\perp}, z) ,  \varphi (t, \vec{x}_{\perp}^{\: \prime}, z^{\prime} ) \right ]
=
-i \: \delta ^{2}(\vec{x}_{\perp} - \vec{x}_{\perp}^{\:\prime} ) \delta (z - z^{\prime})\:,
\label{eq:canonicalcommutaionrelation1}
\\
& & 
\left [ { \varphi } (t, \vec{x}_{\perp}, z) ,  \varphi (t, \vec{x}_{\perp}^{\: \prime}, z^{\prime} ) \right ]
=\left [ \dot{ \varphi } (t, \vec{x}_{\perp}, z) ,  \dot \varphi (t, \vec{x}_{\perp}^{\: \prime}, z^{\prime} ) \right ]
=0\:,
\label{eq:canonicalcommutaionrelation2}
 \end{eqnarray}
 by introducing  commutation relations among some operators which are analogous to the conventional 
 creation and annihilation operators.  Note that  the time derivative is denoted by  dot in (\ref{eq:canonicalcommutaionrelation1}) and (\ref{eq:canonicalcommutaionrelation2}).

The solutions to the equations of motion   (\ref{eq:equationsofmotion})   are given in the form of  a superposition of the fundamental solutions  $\phi_{k}(z; \lambda ) $$ \:\:(k=1,2)$ in (\ref{eq:eigenvalueproblem})  combined  with the plane waves in the $x$- and $y$- directions.  Therefore the scalar field is expanded   generally  as
 \begin{eqnarray}
 \varphi (t, \vec{x}_{\perp}, z) 
 &=&
  \int d\lambda \sum_{k=1, 2} \int \frac{d^{2}\vec{p}_{\perp}}
 {\sqrt{(2\pi)^{2} \: 2E }} \bigg \{
 \alpha_{k}(\vec{p}_{\perp}, \lambda ) e^{i \vec{p}_{\perp} \cdot \vec{x}_{\perp} -iEt} \phi_{k}(z; \lambda)
 \nonumber \\  
 & &
\hskip4.0cm   + 
  \alpha_{k}^{\dag }(\vec{p}_{\perp}, \lambda ) e^{ - i \vec{p}_{\perp} \cdot \vec{x}_{\perp} + iEt}
   \phi_{k}(z; \lambda)
 \bigg \}\:,
 \label{eq:scalaralphacretionannihilation}
 \end{eqnarray}
where $E$ is defined by 
 \begin{eqnarray}
 E=\sqrt{\vec{p}_{\perp}^{\: 2} + \lambda }\:.
 \end{eqnarray}
The coefficients,  $\alpha_{k}( \vec{p}_{\perp} , \lambda )$ and  $\alpha_{k}^{\dag } ( \vec{p}_{\perp} , \lambda )$ $(k=1,2)$  in (\ref{eq:scalaralphacretionannihilation}),  are  regarded as operators, which we postulate to satisfy   the following commutation relations, 
 \begin{eqnarray}
 & & 
\left [  \alpha_{k}(\vec{p}_{\perp}, \lambda ) ,\: 
  \alpha_{l}^{\dag }(\vec{p}_{\perp}^{\: \prime}, \lambda^{\prime} )  \right ]
  =\delta^{2}(\vec{p}_{\perp} - \vec{p}_{\perp}^{\:\prime})\delta(\lambda - \lambda ^{\prime})
  \frac{d\rho_{kl} ( \lambda )}{d\lambda}\:, 
  \label{eq:scalarcommutationrelation1}
 \\
 & &  \left [  \alpha_{k}(\vec{p}_{\perp}, \lambda ) ,\: 
  \alpha_{l}^{ }(\vec{p}_{\perp}^{\: \prime}, \lambda^{\prime} )  \right ]
  =
  \left [  \alpha_{k}^{\dag}(\vec{p}_{\perp}, \lambda ) ,\: 
  \alpha_{l}^{\dag }(\vec{p}_{\perp}^{\: \prime}, \lambda^{\prime} )  \right ]=0\:.
    \label{eq:scalarcommutationrelation2}
 \end{eqnarray}
Note that the spectral density matrix $d\rho_{kl} (\lambda )/d\lambda$ in (\ref{eq:scalarcommutationrelation1}) is symmetric and positive semi-definite.  Employing  (\ref{eq:scalarcommutationrelation1}) and (\ref{eq:scalarcommutationrelation2}) we immediately get 

\begin{eqnarray}
\left [
\dot{\varphi }( t, \vec{x}_{\perp} , z ),  \varphi ( t, \vec{x}_{\perp}^{\:\prime}  , z^{\:\prime} ) \right ] =  -i\: \delta^{2}(\vec{x}_{\perp} - \vec{x}_{\perp}^{\:\prime} )  \sum_{k,l =1,2} \int d\lambda \: \phi_{k}(z, \lambda ) \frac{ d\rho_{kl}(\lambda ) }{d\lambda }  \phi_{l}(z^{\:\prime} ; \lambda )\:,
\end{eqnarray}
and are led to  the canonical commutation relation (\ref{eq:canonicalcommutaionrelation1})  on  using  the completeness  (\ref{eq:completenessrelation0}).  The relation (\ref{eq:canonicalcommutaionrelation2}) can also be derived straightforwardly by using  (\ref{eq:scalarcommutationrelation1}) and (\ref{eq:scalarcommutationrelation2}) .
 
 In case that there does not exist a bubble wall, i.e., $M(z)=M$ (constant),  the operators  $\alpha _{k}^{\dag}  (\vec{p}_{\perp}, \lambda ) $ and  $\alpha _{k} (\vec{p}_{\perp}, \lambda ) $ have to reduce to  the conventional creation and annihilation operators. In fact putting  the solutions (\ref{eq:constantmasseigenfunction}) together with the  spectral functions (\ref{eq:stieltjesmeasure1}),   (\ref{eq:stieltjesmeasure2}) and   (\ref{eq:stieltjesmeasure}), we are able to rewrite (\ref{eq:scalaralphacretionannihilation}) in the following form
 \begin{eqnarray}
 \varphi  (t, \vec{x}_{\perp}, z)
 &=&
 \int_{M^{2}}^{\infty} \frac{d\lambda }{2\sqrt{\lambda - M^{2}}} 
 \int \frac{d^{2}\vec{p}_{\perp}}{\sqrt{(2\pi)^{3} 2 E}}
\bigg \{
\nonumber \\
& & 
\:\:\: a_{-}(\vec{p}_{\perp} , \lambda ) e^{i \vec{p}_{\perp} \cdot \vec{x}_{\perp} + i \sqrt{\lambda - M^{2}} z}
e^{-iEt}  
+a_{+}(\vec{p}_{\perp} , \lambda ) e^{i \vec{p}_{\perp} \cdot \vec{x}_{\perp} - i \sqrt{\lambda - M^{2}} z}
e^{-iEt}  
\nonumber \\
& & 
+ a_{-}^{\dag} (\vec{p}_{\perp} , \lambda ) e^{ - i \vec{p}_{\perp} \cdot \vec{x}_{\perp} - i \sqrt{\lambda - M^{2}} z}e^{ iEt}  
+a_{+}^{\dag}(\vec{p}_{\perp} , \lambda ) e^{ - i \vec{p}_{\perp} \cdot \vec{x}_{\perp} + i \sqrt{\lambda - M^{2}} z}e^{iEt}
\bigg \}  \:, 
\nonumber 
\\
 \end{eqnarray}
 where we have introduced two types of linear combinations of 
 $\alpha_{1}( \vec{p}_{\perp} , \lambda )$    and    $\alpha_{2}( \vec{p}_{\perp} , \lambda )$ , 
 i.e., 
 \begin{eqnarray}
& &
 a_{\pm}(\vec{p}_{\perp} , \lambda ) \equiv \sqrt{2\pi} \left \{
 \sqrt{\lambda - M^{2}}\: \alpha _{1}(\vec{p}_{\perp} , \lambda) \pm i \alpha_{2}(\vec{p}_{\perp}, \lambda ) 
 \right \}\:. 
\end{eqnarray}
The commutation relations among these combinations turn out to be 
\begin{eqnarray}
\left [ a_{+}(\vec{p}_{\perp} , \lambda ), \: a_{+}^{\dag} (\vec{p}_{\perp}^{\:\prime} , \lambda^{\prime} )\right ]
&=&
\delta^{2}(\vec{p}_{\perp} - \vec{p}_{\perp}^{\:\prime} ) \delta ( \sqrt{ \lambda - M^{2}} - 
\sqrt{  \lambda ^{\prime} - M^{2} })\:,
\\
\left [ a_{-}(\vec{p}_{\perp} , \lambda ), \: a_{-}^{\dag} (\vec{p}_{\perp}^{\:\prime} , \lambda^{\prime} )\right ]
&=&
\delta^{2}(\vec{p}_{\perp} - \vec{p}_{\perp}^{\:\prime} ) \delta ( \sqrt{ \lambda - M^{2}} - 
\sqrt{  \lambda ^{\prime} - M^{2} })\:, 
\end{eqnarray}
and all of the other types of commutators    vanish.  Once we identify  $\sqrt{\lambda - M^{2}} $ as the $z$-component of the momentum $\vec{p}$, then it is easy to see that 
$ a_{-}(\vec{p}_{\perp} , \lambda )$ and 
$ a_{+}(\vec{p}_{\perp} , \lambda )$ are the usual annihilation operator of a particle with momentum
$\vec{p}=( \vec{p}_{\perp} , \sqrt{\lambda - M^{2}} )$   and 
$\vec{p}=( \vec{p}_{\perp} , - \sqrt{\lambda - M^{2}} )$,    respectively. 

In the presence of the bubble wall, the interpretation of operators $\alpha_{k}^{\dag} (\vec{p}_{\perp}, \lambda )$ and  $\alpha_{k} (\vec{p}_{\perp}, \lambda )$ is subtle and is not so simple. The notion  of a particle itself is obscured, since the mass is position-dependent.   We may, however,  call 
$\alpha _{k}^{\dag} ( \vec{p}_{\perp} , \lambda )$ and $\alpha _{k} ( \vec{p}_{\perp} , \lambda )$ as
 creation and annihilation operators of some ``quanta'' labelled by $k$, $\vec{p}_{\perp}$ and  $\lambda $.    Suppose that we define the vacuum $\vert 0 \rangle $ by
\begin{eqnarray}
\alpha _{k} ( \vec{p}_{\perp} , \lambda ) \vert 0 \rangle =0, \hskip1cm 
{\rm for}\:\:\: k=1,2,  \:\:\: ^\forall \vec{p}{_\perp} , \:\:\: ^\forall \lambda \:.
\label{eq:vacuum}
\end{eqnarray}
In the symmetry-broken region, the state $\alpha _{k} ^{\dag} ( \vec{p}_{\perp} , \lambda ) \vert 0 \rangle $  
behaves  like a one-particle state  with a definite mass, say, $M$,  and looks like a superposed state  with four-momentum of   $(E, \vec{p}_{\perp} , + \sqrt{\lambda - M^{2}})$ and $( E, \vec{p}_{\perp} , - \sqrt{\lambda - M^{2}})$.  We have to keep in our mind, however, that the normalization of  this state differs from the usual one due to the $\lambda$-dependent quantity  $d\rho_{kl}(\lambda )/d\lambda $  in (\ref{eq:scalarcommutationrelation1}) and that this should be taken into account in  quantitative calculations.  We should also notice that  the two-point function (\ref{eq:integralrepresentation})  is the vacuum expectation value of the time-ordered product of the scalar fields where the vacuum is defined by (\ref{eq:vacuum}).

 \subsection{The Dirac field}
 
 Let us now turn our attention to the Dirac field,  whose general solution to the free equations of motion (\ref{eq:eigenvalueproblemoriginal}) is given by a superposition of (\ref{eq:1901}). We express the solution by separating the part belonging to  the positive ($E>0$) and  negative ($E<0$) energies   in the following way, i.e.,
\begin{eqnarray}
   \psi  (t, \vec{x}_{\perp},  z ) 
   &=&
     \sum _{k=1, 2}\:\:\sum_{s=+, -} \:\: \bigg \{
   \int_{E > 0} dE  \int \frac{d^{2}\vec{p}_{\:\perp} }{2\pi}\: 
   \beta_{k}^{(s)} (\vec{p}_{\perp} , E ) \: \psi_{k}^{(s)} ( z; E )
   e^{  i \vec{p}_{\perp} \cdot \vec{x}_{\perp} - iEt}     
   \nonumber \\
& &   + \int_{E < 0} dE  \int \frac{d^{2}\vec{p}_{\perp} }{2\pi} \gamma_{k}^{(s) \dag}(\vec{p}_{\perp}, E)\psi _{k}^{(s)}(z; E)e^{  i \vec{p}_{\perp} \cdot \vec{x}_{\perp} - iEt}   \bigg \}.    
\end{eqnarray}  
The coefficients $\beta_{k}^{(s)} (\vec{p}_{\perp} , E )$ and 
$\gamma_{k}^{(s) \: \dag} (\vec{p}_{\perp} , E )$ are operators whose anti-commutation relations are arranged   so that the basic  relations 
\begin{eqnarray}
\big  \{  \psi (t, \vec{x}_{\perp}, z ) ,  \psi^{ \: \dag} (t, \vec{x}^{\:\prime} _{\perp}, z^{\:\prime}  )  \big \}
&=&\delta^{2}(\vec{x}_{\perp} - \vec{x}_{\perp}^{\:\prime}  ) \delta (z - z^{\:\prime} )\: , 
\label{eq:psipsianticommutator1}
\\
\big \{  \psi (t, \vec{x}_{\perp}, z ) ,  \psi^{ } (t, \vec{x}^{\:\prime} _{\perp}, z^{\:\prime}  )  \big \} 
&=&0  \: , 
\label{eq:psipsianticommutator2}
\\
\big \{  \psi^{\:\dag}  (t, \vec{x}_{\perp}, z ) ,  \psi^{ \: \dag} (t, \vec{x}^{\:\prime} _{\perp}, z^{\:\prime}  )  
\big \} 
&=&0  \: , 
\label{eq:psipsianticommutator3}
\end{eqnarray}
are realized.   Here we would like to point out that once we postulate the following anti-commutation relations
\begin{eqnarray}
\left \{  \beta_{k}^{(r)} (\vec{p}_{\perp} , E ),  \: \beta_{l}^{(s)\: \dag } (\vec{p}_{\perp}^{\:\prime} , E^{\:\prime}  )  \right \}
=\delta^{rs} \: \delta^{2}( \vec{p}_{\perp} - \vec{p}_{\perp}^{\:\prime} ) \: \delta (E - E^{\: \prime})
\: \frac{d\rho_{kl}^{(s)}(E)}{dE}\:,
\label{betabetaanticommutator}
\end{eqnarray}
\begin{eqnarray}
\left \{  \gamma_{k}^{(r)} (\vec{p}_{\perp} , E ),  \: \gamma_{l}^{(s)\: \dag } (\vec{p}_{\perp}^{\:\prime} , E^{\:\prime}  )  \right \}
=\delta^{rs} \: \delta^{2}( \vec{p}_{\perp} - \vec{p}_{\perp}^{\:\prime} ) \: \delta (E - E^{\: \prime})
\: \frac{d\rho_{kl}^{(s)}(E)}{dE}\:,
\label{eq:gammagammaaniticomutator}
\end{eqnarray}
and vanishing of  all  other types of anti-commutators, then (\ref{eq:psipsianticommutator1}), 
 (\ref{eq:psipsianticommutator2})  and    (\ref{eq:psipsianticommutator3}) are  all satisfied.

As in the scalar case we may  call $\beta _{k}^{(r) \dag} ( \vec{p}_{\perp} , E )$ and $\beta _{k}^{(r)} ( \vec{p}_{\perp} , E )$ as creation and annihilation operators of  Fermionic ``quanta'' labelled by $k$, $r$, $\vec{p}_{\perp}$ and  $E$.    Likewise we may call  $\gamma _{k}^{(r) \dag} ( \vec{p}_{\perp} , E )$ and $\gamma _{k}^{(r)} ( \vec{p}_{\perp} , E )$ as those  of anti-Fermionic ``quanta''  labelled by $k$, $r$, $\vec{p}_{\perp}$ and  $E$.   
We define the vacuum by the conditions 
\begin{eqnarray}
\beta_{k}^{(r)} ( \vec{p}_{\perp}, E) \vert 0 \rangle =0, \:\:\:
\gamma_{k}^{(r)} ( \vec{p}_{\perp}, E) \vert 0 \rangle =0, \:\:\:{\rm for} \:\:\:
k=1,2, \:\: r=\pm, \:\: ^\forall \vec{p}_{\perp}, \:\:  ^\forall E\:.
\end{eqnarray}
Then it turns out that the Fermion's propagator (\ref{eq:fermionpropagatorfinalform}) supplemented by  the $+i\varepsilon $ prescriptions (\ref{eq:prescription1}) and (\ref{eq:prescription2}) is an expectation value of the  time-ordered product  of Fermion's bilinear  operators with respect to this vacuum. 
In the symmetry-broken region, the states $\beta_{k} ^{(r) \dag} ( \vec{p}_{\perp} , E ) \vert 0 \rangle $  and $\gamma_{k} ^{(r) \dag} ( - \vec{p}_{\perp} , E ) \vert 0 \rangle $ 
behave  like a one-particle Fermion and anti-Fermion states, respectively,   with a definite mass, say, $m$ and are a superposition of  the  four-momentum states of   
$(\vert E \vert, \vec{p}_{\perp} , +q)$ and 
$(\vert E \vert, \vec{p}_{\perp} , -q)$, as a linear combination of  $k=1$ and $k=2$  states.
Here $q$ is given by  
\begin{eqnarray}
q=\sqrt{E^{2}-(m^{2} +\vert \vec{p}_{\perp} \vert^{2} )}.
\end{eqnarray}

In the absence of the bubble wall,  we have $m(z)=m$ (constant) and 
$\beta_{k}^{(s)} (\vec{p}_{\perp} , E )$ and $\gamma_{k}^{(s) } (\vec{p}_{\perp} , E )$  may be shown to go over to the usual annihilation operators. In fact 
 \begin{eqnarray}
 b^{(\pm)}(\vec{p}_{\perp} , q )
 &=&
 \kappa^{(\pm )} \beta_{1}^{(\pm )} (\vec{p}_{\perp} , E ) + i \kappa^{(\mp)} \beta_{2}^{(\pm )}
 (\vec{p}_{\perp} , E)\:,
 \\
  b^{(\pm)}(\vec{p}_{\perp} , - q )
  &=&
   \kappa^{(\pm )} \beta_{1}^{(\pm )} (\vec{p}_{\perp} , E ) - i \kappa^{(\mp)} \beta_{2}^{(\pm )}
 (\vec{p}_{\perp} , E)\:,
 \end{eqnarray}
  are the annihilation operators of the Fermion with the momentum $\vec{p}=(\vec{p}_{\perp} , q) $ and 
  $\vec{p}=(\vec{p}_{\perp} , - q) $, respectively, where  we defined
   \begin{eqnarray}
 \kappa ^{(\pm)}=\sqrt{2 \pi } \sqrt{ \frac{\vert E \vert \pm \sqrt{m^{2} + \vert \vec{p}_{\perp} 
 \vert ^{2} }}{2 \vert E \vert } }\:.
 \end{eqnarray}
We are able to confirm  the usual anti-commutators
\begin{eqnarray}
\left \{ b^{(\pm)} (\vec{p}_{\perp} , q ),   b^{(\pm)\:\dag } (\vec{p}^{\:\prime} _{\perp} , q^{\:\prime} )\right \}
= \delta ^{2} (\vec{p}_{\perp} - \vec{p}_{\perp}^{\:\prime} )\delta( q - q^{\: \prime})\:,
\\
\left \{ b^{(\pm)} (\vec{p}_{\perp} , -q ),   b^{(\pm)\:\dag } (\vec{p}^{\:\prime} _{\perp} , -q^{\:\prime} )\right \}
= \delta ^{2} (\vec{p}_{\perp} - \vec{p}_{\perp}^{\:\prime} )\delta( q - q^{\: \prime})\;,
\end{eqnarray} 
 and  all other types anti-commutators vanish.

Similarly, the creation operators of the anti-Fermion with momentum $\vec{p}=(\vec{p}_{\perp} , q) $ and 
  $\vec{p}=(\vec{p}_{\perp} , - q) $ are expressed,  respectively,  by  
 \begin{eqnarray}
 c^{(\pm)\: \dag }( - \vec{p}_{\perp} , - q )
 &=&
  \kappa^{(\pm )} \gamma_{1}^{(\mp ) \: \dag } (\vec{p}_{\perp} , E ) - i \kappa^{(\mp)}
   \gamma_{2}^{(\mp )\: \dag } (\vec{p}_{\perp} , E)\:,
 \\
  c^{(\pm)\: \dag }( - \vec{p}_{\perp} ,  q )
  &=&
    \kappa^{(\pm )} \gamma_{1}^{(\mp ) \: \dag } (\vec{p}_{\perp} , E ) + i \kappa^{(\mp)}
   \gamma_{2}^{(\mp )\: \dag } (\vec{p}_{\perp} , E)\:, 
 \end{eqnarray}
 and are shown to satisfy the anti-commutators 
\begin{eqnarray}
\left \{ c^{(\pm)} (- \vec{p}_{\perp} , - q ),   c^{(\pm)\:\dag } ( -\vec{p}^{\:\prime} _{\perp} , - q^{\:\prime} )\right \}
= \delta ^{2} (\vec{p}_{\perp} - \vec{p}_{\perp}^{\:\prime} )\delta( q - q^{\: \prime})\:,
\\
\left \{ c^{(\pm)} (-\vec{p}_{\perp} , q ),   c^{(\pm)\:\dag } (-\vec{p}^{\:\prime} _{\perp} , q^{\:\prime} )\right \}
= \delta ^{2} (\vec{p}_{\perp} - \vec{p}_{\perp}^{\:\prime} )\delta( q - q^{\: \prime})\;,
\end{eqnarray} 
together with vanishing  of  all other types of anti-commutators.

\section{Concluding Remarks}

In the presence of the Higgs  condensate bubble,  we have successfully constructed the two-point Green's functions of scalar, spinor and vector fields in the form of integral representations. The key role of the construction is played by  the spectral functions, which we are in principle able to compute  provided that the large $\vert z \vert$ behavior of fundamental  solutions are known. The spectral functions also tell us about  the range of the spectra and therefore we can identify   the infrared region simply by looking at the integral representation.  With the help of these propagators, we are able to check the interplay  between virtual and real soft emitted gauge bosons,  and natures of cancellation of infrared singularities. 

The Sudakov-type double logarithms and variations thereof may also be investigated by using our Green's functions. As we see in (\ref{eq:fermionpropagatorfinalform}),  the denominators of the Fermion's propagator look like those in the old-fashioned perturbation.   Of course the denominators of the Bosonic propagator  (\ref{eq:integralrepresentation}) may also be divided into two to  cast them into the form of old-fashioned perturbation. It has been known that,  in the  very high energy limit, only a few  among many terms in the old-fashioned perturbation survive the limit and perturbative calculation rules become very much simplified \cite{weinberg1}.  It has also been known that the derivation of the Sudakov-type double logarithms and their all-order resummation  are possible by picking up only limited number of terms in the old-fashioned perturbation theory \cite{jackiw}. The propagators that we derived in the present paper may  be adapted   to such  analyses. 

For more dedicated analyses on the dynamics near the bubble wall interface, finite temperature Green's functions must be worked out, which, however, we will report in separate publications.  In constructing finite temperature Green's functions, we have to take care of the operator ordering. The commutation and anti-commutation relations that we have postulated in Sect.  \ref{sec:fieldquantization} may  provide us with a good testing-ground of their usefulness.

\acknowledgments
The author would like to thank Drs. J. Ho, O-K. Kwon, S.-A Park and S.-H. Yi for calling his attention to \cite{ho1} and \cite{ho2}.

\paragraph{Note added:} 
After submitting the present paper for publication, two papers,  \cite{ho1} and \cite{ho2},  came to the author's attention.  In these papers another  new approach to field quantization under certain backgrounds is developed.



\begin{thebibliography}{99}

\bibitem{abbott0}
B.P. Abbott et al.,
``Observation of Gravitational Waves from a Binary Black Hole Merger''
Phys. Rev. Letters 116, 061102 (2016),    arXiv:1602.03837 [gr-qc].

DOI: https://doi.org/10.1103/PhysRevLett.116.061102
%
%
\bibitem{abbott1}
B.P. Abbott et al.,
``GW151226: Observation of Gravitational Waves from a 22-Solar-Mass Binary Black Hole  Coalescence''
Phys. Rev. Letters 116, 241103 (2016),    arXiv:1606.04855 [gr-qc].

DOI: https://doi.org/10.1103/PhysRevLett.116.241103
%
%
\bibitem{abbott2}
B.P. Abbott et al.,  
``GW170104: Observation of a 50-Solar-Mass Binary Black Hole Coalescence at Redshift 0.2''
Phys. Rev. Letters 118, 221101 (2017),     arXiv:1706.01812 [gr-qc].

DOI: https://doi.org/10.1103/PhysRevLett.118.221101
%
\bibitem{abbott3}
B.P. Abbott et al.,  
``GW170817: Observation of Gravitational Waves from a Binary Neutron Star Inspiral''
Phys. Rev. Letters 119, 161101 (2017),    arXiv:1710.05832 [gr-qc].

DOI: https://doi.org/10.1103/PhysRevLett.119.161101
%
%
\bibitem{witten}
E. Witten,  ``Cosmic separation of phases'' Phys. Rev. D 30 (1984)  272 - 285.

DOI: https://doi.org/10.1103/PhysRevD.30.272
%
%
\bibitem{hogan}
C. J. Hogan, 
``Gravitational radiation from cosmological phase transitions'',
Mon. Not. Roy. Astron. Soc. 218 (1986) 629 - 636.

DOI: https://doi.org/10.1093/mnras/218.4.629
%
\bibitem{kosowsky1}
A. Kosowsky, M.S. Turner and R. Watkins, 
``Gravitational waves from first order cosmological phase transitions''
Phys. Rev. Letters 69 (1992) 2026.

DOI:  https://doi.org/10.1103/PhysRevLett.69.2026
%
\bibitem{kosowsky2}
M. Kamionkowski, A. Kosowsky, and M.S. Turner,
``Gravitational radiation from first order phase transitions''
Phys. Rev. D 49 (1994) 2837 - 2851.

DOI: https://doi.org/10.1103/PhysRevD.49.2837

astro-ph/9310044
%
\bibitem{kajantie}
K. Kajantie, M. Laine, K. Rummukainen and M.E. Shaposhnikov,    
``Is There a Hot Electroweak Phase Transition at $m_{H} \geq m_{W} $ ?''
Phys. Rev. Letters 77 (1996) 2887.

DOI: https://doi/org/10.1103/PhysRevLett.77.2887
%
\bibitem{karsch}
F. Karsch, T. Neuhaus, A. Patkos and J. Rank,   ``Critical Higgs Mass and Temperature Dependence of Gauge Boson Masses in the $SU(2)$ Gauge-Higgs Model''

Nucl. Phys. B Proc. Suppl. 53 (1997) 623 - 625.

DOI: https://doi.org/10.1016/S0920-5632(96)00736-0 
%
\bibitem{gurtler}
M. Gurtler, E.-M. Ilgenfritz and A. Schiller, ``Where the electroweak phase transition ends''

Phys. Rev. D 56 (1997) 3888.

DOI:   https://doi.org/10.1103/PhysRevD.56.3888
%
\bibitem{rummukainen}
K. Rummukainen, M. Tsypin, K. Kajantie, M. Laine and M.E. Shaposhnikov, ``The Universality Class of the Electroweak Theory'', Nucl. Phys. B 532 (1998) 283 - 314. 

DOI: https://doi.org/10.1016/S0550-3213(98)00494-5
%
\bibitem{csikor}
F. Csikor, Z. Fodor and J. Heitger, ``The Strength of the Electroweak Phase Transition at $m(H) \approx 80$  GeV''  
Phys. Letters B441 (1999) 354 - 362.

DOI: https://doi.org/10.1016/S0370-2693(98)01127-7

%
\bibitem{aoki}
Y. Aoki, F. Csikor, Z. Fodor and A. Ukawa, ``The end point of the first-order phase transition of the $SU(2)$ gauge-Higgs model on a 4-dimensional isotropic lattice'', 
Phys. Rev. D 60 (1999) 0130001.

DOI:   https://doi.org/10.1103/PhysRevD.60.013001
%
\bibitem{laine}
M. Laine and K. Rummukainen, 
Nucl. Phys. B Proc. Suppl. 73 1999 180 - 185, 
``What's new with the electroweak phase transition?''
[hep-lat/9809045].

DOI: https://doi.org/10.1016/S0920-5632(99)85017-8
%
\bibitem{donofrio}
M. D'Onofrio and K. Rummukainen, ``Standard model cross-over on the lattice''  Phys. Rev D 93 (2016) 025003,  arXiv:1508.07161 1

DOI:  https://doi.org/10.1103/PhysRevD93.025003
%
\bibitem{olivergould}
O. Gould, S. G{\" u}yer and   K. Rummukainen, ``First-order electroweak phase transitions: A nonperturbative update'',  Phys. Rev. D 106(2022) 114507, arXiv:2205.07238 [hep-lat].

DOI: 10.1103/PhysRevD.106.114507
%
\bibitem{anderson}
G.W. Anderson and L.J. Hall, ``The electroweak phase transition and baryogenesis''
Phys. Rev. D 45 (1992) 2685.

DOI:   https://doi.org/10.1103/PhysRevD.45.2685
%
\bibitem{quiros}
M. Quir{\' o}s, ``Finite temperature field theory and phase transitions''
ICTP summer school in high-energy physics and cosmology (1999) p. 187 

hep-ph/9901312
%
\bibitem{grojean}
C. Grojean, G. Servant and J.D. Wells, ``First-order electroweak phase transition in the standard model
with a low cutoff''    Phys. Rev. D 71 (2005) 036001,  hep-th/0407019.

DOI:   https://doi.org/10.1103/PhysRevD.71.036001
%
\bibitem{caprini1}
C. Caprini et al, ``Science with the space-based interferometer eLISA. II: gravitational waves from cosmological phase transitions''
JCAP 04 (2016) 001,   arXiv: 1512.06239 [astro-ph.CO]

DOI: 10.1088/1475-7516/2016/04/001
%
\bibitem{caprini2}
C. Caprini et al, ``Detecting gravitational waves from cosmological phase transitions with LISA: 
an update'', JCAP 03 (2020) 024,   arXiv: 1910.13125 [astro-ph.CO]

DOI: 10.1088/1475-7516/2020/03/024
%
\bibitem{hindmarshx}
M. Hindmarsh, M. Luben, J. Lumma and M. Pauly, ``Phase transitions in the early universe" 
SciPost Physics Lecture Notes, 24 (2021) 1-61, arXiv:2008.09136 [astro-ph.CO]

doi:10.21468/SciPostPhysLecNotes.24
%
\bibitem{auclair}
LISA Cosmology Working Group, P. Auclair et al., ``Cosmology with the Laser Interferometer Space Antenna'',   Living Rev.   Rel.  26 (2023) 1 - 254,    arXiv: 2204.05434 [astro-ph.CO].

https://doi.org/10.1007/s41114-023-00045-2  
%
\bibitem{kawamura}
S. Kawamura et al, ``Space gravitational-wave antennas DECIGO and B-DECIGO'' , Int. J. Mod. Phys. D 28 (2019) 12, 1845001.

doi:101142/S0218271818450013
%
\bibitem{ruan}
W.-H. Ruan, Z.-K. Guo, R.-G.Cai and Y.-Z. Zhang,   ``Taiji program: Gravitational-wave sources'', Int. J. Mod. Phys. A 35 (2020) 17, 2050075, arXiv:1807.09495 [gr-qc].

doi:101142/S0217751X2050075X
%
\bibitem{luo}
J. Luo et al., ``TianQin: a space-borne gravitational wave detector'', Class. Quantum Grav. 33 (2016) 3, 035010, arXiv:1512.02076 [astro-ph.IM]. 

doi: 101088/0264-9381/33/3/035010
%
\bibitem{kuzmin}
V.A. Kuzmin, V.A. Rubakov and M.E. Shaposhnikov, 
``On the Anomalous Electroweak Baryon Number Nonconservation in the Early Universe'' 
Phys. Lett B 155 (1985) 36 - 42.

https://doi.org/10.1016/0370-2693(85)91028-7


%
\bibitem{shaposhnikov}
M.E. Shaposhnikov, 
``Baryon Asymmetry of the Universe in Standard Electroweak Theory''
Nucl. Phys. B 287 (1987) 757- 775.

https://doi.org/10.1016/0550-3213(87)90127-1
%
\bibitem{cohen1}
A.G. Cohen, D.B. Kaplan and A.E. Nelson, 
``Progress in Electroweak Baryogenesis'',    
Annu. Rev. Nucl. Part. Sci., 43 (1993) 27 - 70, hep-ph/9302210 [hep-ph] .  

https://doi.org/10.1146/annurev.ns.43.120193.000331  

%
%
\bibitem{konstandin}
T. Konstandin, 
``Quantum Transport and Electroweak Baryogenesis'' 
Phys. Usp. 56 (2013) 747,   arXiv: 1302.6713 [hep-ph].

https://doi.org/10.3367/ufne.0183.201308a.0785
%
\bibitem{morrissey}
D.E. Morrissey and M.J. Ramsey-Musolf, 
``Electroweak Baryogenesis'' 
New J. Phys. 14 (2012) 125003, arXiv: 1206.2942 [hep-ph]

DOI:10.1088/1367-2630/14/12/125003
%
\bibitem{garbrecht}
B. Garbrecht, 
``Why is there more matter than antimatter ?'' , 
Progr. Part.  Nucl. Phys. 110 (2020) 103727  1-46,  arXiv: 1812.02651 [hep-ph]


https://doi.org/10.1016/j.ppnp.2019.103727
%
%
%
%
%
\bibitem{boedeker1}
D. B{\" o}deker and G.D. Moore, 
``Can electroweak bubble walls  run away?'', 
JCAP 05 (2009) 09,  arXiv: 0903.4099 [hep-ph].

DOI: 10.1088/1475-7516/2009/05/009
%
%
\bibitem{espinosa1}
J.R. Espinoza, T. Konstandin, J.M. No and G. Servant,  ``Energy budget of cosmological first-order phase transitions''  JCAP  06 (2010) 028,  arXiv: 1004.4187 [hep-ph].

https://doi.org/10.1088/1475-7516/2010/06/028

%
%
%
\bibitem{boedeker2}
D. B{\" o}deker and G.D. Moore, 
`` Electroweak bubble wall speed limit'',
JCAP 05 (2017) 025,  arXiv: 1703.08215 [hep-ph]

DOI: 10.1088/1475-7516/2017/05/025
%
%
\bibitem{ginzburgfrank}
V.L. Ginzburg and I.M. Frank, 
``Radiation of a uniformly moving electron due to its transition from one medium into another''
J. Phys. (USSR) 9 (1945) 353 - 362; Zh. Eksp. Teor. Fiz. 16 (1946) 15-28.
%
\bibitem{jackson}
J.D. Jackson, ``Classical Electrodynamics'' (Wiley, 1998)

ISBN-10:9780471309321
%
%
\bibitem{hoeche}
S. H{\" o}che, J. Kozaczuk, A.J. Long, J. Turner and Y. Wang, 
``Towards an all-orders calculation of the electroweak bubble wall velocity'', 
JCAP  03 (2021) 009,   arXiv: 2007.10343 [hep-ph].

DOI: 10.1088/1475-7516/2021/03/009 
%
%
%
%
%
\bibitem{sudakov}
V.V. Sudakov, 
``Vertex Parts at Very High Energies in Quantum Electrodynamics''
Soviet Physics JETP 3 (1956) 65 - 71.
%
%
%
\bibitem{mueller}
A.H. Mueller, ``Asymptotic behavior of the Sudakov form factor''
Phys. Rev. D20 (1979)  2037 - 2046.


DOI: https://doi.org/10.1103/PhysRevD.20.2037
%
\bibitem{collins3}
J.C. Collins, ``Algorithm to compute corrections to the Sudakov form factor''
Phys. Rev. D22 (1980)  1478 - 1489.


DOI: https://doi.org/10.1103/PhysRevD.22.1478
%
\bibitem{sen1}
A. Sen, 
``Asymptotic behavior of the Sudakov form factor in quantum chromodynamics''
Phys. Rev. D24 (1981) 3281 - 3304. 


DOI: https://doi.org/10.1103/PhysRevD.24.3281
%
\bibitem{sen2}
A. Sen, 
``Asymptotic behavior of the fixed-angle on-shell scattering amplitudes in non-Abelian gauge theories''
Phys. Rev. D28 (1983) 860 - 875. 


DOI: https://doi.org/10.1103/PhysRevD.28.860
%
%
\bibitem{collins2}
J.C. Collins, 
``Sudakov Form Factor''
in ``Perturbative QCD'' (ed. by A.H. Mueller) Advanced Series on  Directions in  High Energy Physics 
 {\rm 5}, 1 (1988, ISBN: 978-981-4503-26-6)
%
\bibitem{contopanagos}
H. Contopanagos, E. Laenen, and G.F. Sterman,
``Sudakov factorization and resummation'' 
Nucl. Phys. B 484 (1997) 303 - 330,   hep-ph/9604313

https://doi.org/10.1016/50550-3213(96)00567-6
%
%
%
%
%
\bibitem{marc}
M.B. Mancha, T. Prokopec and B. {\' S}wi{\` e}zewska,  ``Field-theoretic derivation of bubble-wall force''
JHEP 01 (2021) 070, arXiv:2005.10875 [hep-th]

DOI: https://doi.org/10.1007/JHEP01(2021)070  
%
\bibitem{buckley}
A. Buckley et al., 
``General-purpose event generators for  LHC physics'',
Physics Reports 504 (2011) 145-233, arXiv:1101.2599 [hep-ph].

doi: 10.1016/j.physrep.2011.03.005
%
\bibitem{webber}
B.R. Webber, ``Monte Carlo Simulation of Hard Hadronic Processes''
Ann. Rev. Nucl. Part. Sci. 36 (1986) 253 - 286, arXiv: hep-ph/0312336

https://doi.org/10.1146/annurev.ns.36.120186.001345
%
%
%
\bibitem{dennerpozzorini1}
A. Denner and S. Pozzorini, 
``One-loop leading logarithms in electroweak radiative corrections I. Results''
Eur.   Phys.    J.   C 18 (2001) 461- 480, arXiv:hep-ph/0010201

https://doi.org/10.1007/s100520100551
%
\bibitem{dennerpozzorini2}
A. Denner and S. Pozzorini, 
``One-loop leading logarithms in electroweak radiative corrections II. Factorization of collinear singularities''

Eur.   Phys.    J.   C 21 (2001) 63-79, arXiv:hep-ph/0104127

https://doi.org/10.1007/s100520100721
%
\bibitem{dennerpozzorini3}
A. Denner and S. Pozzorini, 
``An algorithm for the high-energy expansion of multi-loop diagrams to next-to-leading logarithmic accuracy''
Nucl. Phys. B717 (2005) 48-85, arXiv:hep-ph/0408068

https://doi.org/10.1016/j.nuclphysb.2005.03.036
%
\bibitem{rothdenner}
M. Roth and A. Denner, 
``High-Energy Approximation of One-Loop Feynman Integrals''
Nucl. Phys. B 479 (1996) 495 - 514,  hep-ph/9605420.

https://doi.org/10.1016/0550-3213(96)00435-X
%
%
\bibitem{fadin}
V.S. Fadin, L.N. Lipatov, A.D. Martin and M. Melles, 
`` Resummation of double logarithms in electroweak high energy processes'',  
Phys. Rev. D 61 (2000) 094002  1-13,    arXiv:hep-ph/9910338 

DOI: https://doi.org/10.1103/PhysRevD.61.094002
%
\bibitem{meles}
M. Melles, ``Subleading Sudakov logarithms in electroweak high energy processes to all orders''
Phys. Rev.  D 63 (2001) 034003, arXiv:hep-ph/0004056

DOI: https://doi.org/10.1103/PhysRevD.63.034003
%
\bibitem{pagani}
D. Pagani and M. Zaro, ``One-loop electroweak Sudakov logarithms: a revisitation and automation''
JHEP 02 (2022) 161,   arXiv: 2110.03714 [hep-ph]

https://doi.org/10.1007/JHEP02(2022)161
%
\bibitem{lindert}
J.M. Lindert and L. Mai, ``Logarithmic EW corrections at one-loop''

arXiv: 2312.07927 [hep-ph]

%
\bibitem{laurent1}
B. Laurent and J.M. Cline, ``Fluid equations for fast-moving electroweak walls''
Phys. Rev. D 102 (2020) 063516,  arXiv:2007.10935 [hep-ph]

DOI: https://doi.org/10.1103/PhysRevD.102.063516
%
\bibitem{laurent2}
B. Laurent and J.M. Cline, ``First principles determination of bubble wall velocity''
Phys. Rev. D 106 (2022) 023501, arXiv:2204.13120 [hep-ph].

DOI: https://doi.org/10.1103/PhysRevD.106.023501
%
\bibitem{curtis1}
S. De Curtis, L.Delle Rose, A. Guiggiani, A.G. Muyor and G. Panico, ``Bubble wall dynamics at the electroweak phase transition''  JHEP 03 (2022) 163, arXiv:2201.08220 [hep-ph]

https://doi.org/10.1007/JHEP03(2022)163
%
\bibitem{curtis2}
S. De Curtis, L.Delle Rose, A. Guiggiani, A.G. Muyor and G. Panico, ``Collision Integrals for Cosmological Phase Transitions'' JHEP 05 (2023) 194, arXiv:2303.05846 [hep-ph]

https://doi.org/10.1007/JHEP05(2023)194
%
\bibitem{curtis3}
S. De Curtis, L.Delle Rose, A. Guiggiani, A.G. Muyor and G. Panico, ``Non-linearities in cosmological bubble wall dynamics''
arXiv:2401.13522 [hep-ph]

%
%
\bibitem{balaji}
S. Balaji, M. Spannowsky and C. Tamarit, 
``Cosmological bubble friction in local equilibrium'' JCAP 03 (2021) 051,  arXiv:2010.08013 [hep-ph].

DOI: 10.1088/1475-7516/2021/03/051
%
\bibitem{ai1}
W.-Y.Ai, B. Garbrecht and C. Tamarit,  ``Bubble wall velocities in local equilibrium''
JCAP 03 (2022) 015,   arXiv: 2109.13710 [hep-ph]

DOI: 10.1088/1475-7516/2022/03/015
%
\bibitem{ai2}
W.-Y.Ai, B. Laurent and J. van de Vis,  ``Model-independent bubble wall velocities in local thermal equilibrium'', JCAP 07 (2023) 002,  arXiv: 2303.10171 [astro-ph.CO].

DOI: 10.1088/1475-7516/2023/07/002
%
\bibitem{azatov2}
A. Azatov and M. Vanvlasselaer, ``Phase transitions in perturbative walking dynamics''  JHEP 09 (2020) 085,  arXiv:2003.10265 [hep-ph]

https://doi.org/10.1007/JHEP09(2020)085
%
%
\bibitem{GouttennoireJinnoSala}
Y. Gouttenoire, R. Jinno and F. Sala, ``Friction pressure on relativistic bubble walls'', JHEP 05 (2022) 004, arXiv:2112.07686 [hep-ph]. 

https://doi.org/10.1007/JHEP05(2022)004
%
%
\bibitem{garcia1}
I.G. Garcia, G. Koszegi and R. Petrossian-Byrne,  ``Reflections on Bubble Walls'' JHEP 09 (2023) 013,  arXiv: 2212.10572 [hep-ph]

https://doi.org/10.1007/JHEP09(2023)013
%
%
\bibitem{azatov3}
A. Azatov and M. Vanvlasselaer, ``Bubble wall velocity: heavy physics effects'' JCAP 01 (2021) 058,  arXiv:2010.02590 [hep-ph]

DOI: 10.1088/1475-7516/2021/01/058
%
%
%
\bibitem{dorsch0}
G. C. Dorsch, S.J. Huber and T. Konstandin,  ``On the wall velocity dependence of electroweak baryogenesis''  JCAP 08 (2021) 020,   arXiv:2106.06547     [hep-ph].

DOI: https://doi.org/10.1088/1475-7516/2021/08/020
%
%
%
%
\bibitem{dorsch}
G. C. Dorsch, S.J. Huber and T. Konstandin,  ``A sonic boom in bubble wall friction'' JCAP 04 (2022) 010, 
arXiv:2112.12548 [hep-ph]

DOI: 10.1088/1475-7516/2022/04/010
%
%
\bibitem{dorsch2}
G. C. Dorsch   and D.A. Pinto,  ``Bubble wall velocities with an extended fluid Ansatz''

arXiv:2312.02354  [hep-ph]


%
%
\bibitem{azatov4}
A. Azatov, G. Barni, R.Petrossian-Byrne and M. Vanvlasselaer, 
``Quantisation Across Bubble Walls and Friction''

arXiv: 2310.06972 [hep-ph]
%
\bibitem{weyl}
H. Weyl, ``{\" U}her gew{\" o}hnliche Differentialgleichungen mit Singularit{\" a}ten und die zugeh{\" o}rigen Entwicklungen willk{\" u}rlicher Funktionen'',  Mathematische Annalen {\bf 68} (1910) 220 - 269.

https://doi.org/10.1007/BF01474161.

\bibitem{stone}
M.H. Stone, ``Linear Transformation in Hilbert Space and their Applications to Analysis'',   American Mathematical Society Colloquium Publication, vol. 15 (New York, 1932)

ISBN: 9780821810156[0821810154]
%
%
\bibitem{titchmarsh}
E.C. Titchmarsh, ``Eigenfunction expansions associated with second-order differential equation''  (Oxford, Clarendon Press,  1946)
%
\bibitem{kodaira1}
K. Kodaira, 
``The Eigenvalue Problem for OrdinaryDifferential Equation of the Second Order and Heisenberg's Theory of S-Matrices''
American Journal of Mathematics, {\bf 71} (1949) 921 - 945.

https://www.jstor.org/stable/2372377
%
   \bibitem{kodaira2}
   K. Kodaira, 
    ``On ordinary differential equations of any even order and corresponding eigenfunction expansion''  American Journal of Mathematics, {\bf 72} (1950) 502 - 544.
%

https://www.jstor.org/stable/2372051

\bibitem{yoshida}
K. Yosida, 
``On Titchmarsh-Kodaira's  formula concerning Weyl-Stone's  eigenfunction expansion''
Nagoya Mathematical  Journal 1  (1950) 49-58.

https://doi.org/10.1017/S0027763000022820
%
\bibitem{yosida2}
K. Yosida, ``Lectures on Differential and Integral Equations'' (Dover Publication, New York, 1991)

ISBN-13: 978-0486666792
%
\bibitem{polyakov1}
A. M. Polyakov, ``Particle spectrum in quantum field theory''
JETP Lett. 20 (1974) 194 - 195.
%
%
\bibitem{ayala}
A. Ayala, J.J-Marlan, L. McLerran and A.P. Vischer, 
``Scattering in the presence of electroweak phase transition bubble wall''
Phys. Rev. D 49 (1994) 5559 - 5570.


DOI: https://doi.org/10.1103/PhysRevD.49.5559
%
%
%
\bibitem{farrar}
G.R. Farrar and J.W. McIntosh, 
``Scattering from a domain wall in a spontaneously broken gauge theory''
Phys. Rev. D 51 (1995) 5889 - 5904,   arXiv:hep-ph/9412270. 


DOI: https://doi.org/10.1103/PhysRevD.51.5889
%
%
%
%
\bibitem{conte}
S.D. Conte and W.C. Sangren, 
``An Expansion Theorem for a Pair of  Singular First Order Equations''
Canadian Journal of Mathematics,  6 (1954) 554 - 560.

https://doi.org/10.4153/CJM-1954-060-0
%
\bibitem{roos}
B.W. Roos and W.C. Sangren, ``Spectra for a Pair of Singular First Order Differential 
Equations''  Proceedings of the American Mathematical Society, 
12 (1961) 468 - 476.

https://www.jstor.org/stable/2034220
%
\bibitem{titchmarsh2}
E.C. Titchmarsh, 
`` Some Eigenfunction Expansion Formulae''
Proceedings of the  London Mathematical Society  (3) 11 (1961) 159-168.

 https://doi.org/10.1112/plms/s3-11.1.159
%
\bibitem{levitan2}
B.M. Levitan and I.S. Sargsjan, 
``Introduction to Spectral Theory: Selfadjoint Ordinary Differential Operators''
(Translations of Mathematical Monographs, Vol. 39, American Mathematical Society, 1975)

ISBN O-8218-1589-X
%
\bibitem{levitan}
B.M. Levitan and I.S. Sargsjan, 
``Sturm-Liouville and Dirac Operators'' (
Kluwer Academic Publishers,  1991) .

ISBN 978-94-010-5667-0
%
%
\bibitem{weinberg1}
S. Weinberg, ``Dynamics at Infinite Momentum'', 
Phys. Rev. 150 (1966) 1313 - 1318.


DOI: https://doi.org/10.1103/PhysRev.150.1313
%
\bibitem{jackiw}
R. Jackiw, ``Dynamics at High Momentum and the Vertex Function of Spinor Electrodynamics'', 
Annals of Physics 48 (1968) 292 - 321.

https://doi.org/10.1016/0003-4916(68)90087-0
%
\bibitem{ho1}
O-K. Kwon, J. Ho, S.-A Park and S.-H. Yi, 
``Toward Quantization of Inhomogeneous Field Theory'', 
Eur. Phys. J. Plus 138 (2023) 202.

https://doi.org/10.1140/epjp/s13360-023-03822-8

arXiv:2206.13210 [hep-th] 
%
\bibitem{ho2}
J. Ho, O-K. Kwon, S.-A Park and S.-H. Yi, 
``Supersymmetric Backgrounds in $(1+1)$ Dimensions and Inhomogeneous Field Theory'', 
JHEP 11 (2023) 219.

https://doi.org/10.1007/JHEP11(2023)219

arXiv:2211.05699 [hep-th]
%

\end{thebibliography}
\end{document}